\documentclass[submission, Phys]{SciPost}

\usepackage[english]{babel}
\usepackage{geometry}
\usepackage{tabularx}
\usepackage{setspace}
\usepackage{blindtext}
\usepackage{graphicx}
\usepackage{abstract}
\usepackage{amsmath}
\usepackage{mathtools}
\usepackage{amsfonts,amssymb}
\usepackage{siunitx}
\usepackage{braket}
\usepackage{xcolor}
\usepackage{pgf}
\usepackage{graphicx}
\usepackage{placeins}
\usepackage{comment}
\usepackage{subcaption}
\usepackage{csquotes}
\usepackage[backend=biber,style=numeric-comp,sorting=none]{biblatex}
\usepackage{hyperref}
\usepackage{orcidlink}
\hypersetup{colorlinks=true,allcolors=blue}
\usepackage{bm}
\usepackage{xcolor}

\hypersetup{
    colorlinks,
    linkcolor={red!50!black},
    citecolor={blue!50!black},
    urlcolor={blue!80!black}
}

\usepackage[bitstream-charter]{mathdesign}
\DeclareSymbolFont{usualmathcal}{OMS}{cmsy}{m}{n}
\DeclareSymbolFontAlphabet{\mathcal}{usualmathcal}

\begin{document}

\begin{center}{\Large \textbf{
Feedback cooling of fermionic atoms in optical lattices\\
}}\end{center}

\begin{center}

Wenhua~Zhao\,\orcidlink{0009-0004-5721-607X}$^1$$^,$$^2$,
Ling-Na Wu\orcidlink{0000-0003-4722-5883}$^1$$^,$$^3$,
Francesco Petiziol\orcidlink{0000-0001-8412-0509}$^1$ and  Andr\'e Eckardt$^1$
            
\end{center}

\begin{center}

$^1$Institut f{\"u}r Theoretische Physik, Technische Universit{\" a}t Berlin, Hardenbergstraße 36, 10623 Berlin, Germany\\
$^2$Max-Born-Institut, 12489 Berlin, Germany\\
$^3$Center for Theoretical Physics \& School of Physics and Optoelectronic Engineering, Hainan University, Haikou, Hainan 570228, China

${}^\star${\small \sf wenhua.zhao@physik.hu-berlin.de,lingna.wu@hainanu.edu.cn, f.petiziol@tu-berlin.de, eckardt@tu-berlin.de}

\end{center}

\begin{center}
\today
\end{center}

\section*{Abstract}
{\bf

We discuss the preparation of topological insulator states with fermionic ultracold atoms in optical lattices by means of measurement-based Markovian feedback control. The designed measurement and feedback operators induce an effective dissipative channel that stabilizes the desired insulator state, either in an exact way or approximately in the case where additional experimental constraints are assumed. 
Successful state preparation is demonstrated in one-dimensional insulators as well as for Haldane's Chern insulator, by calculating the fidelity between the target ground state and the steady state of the feedback-modified master equation. The fidelity is obtained numerically through exact diagonalization or via time evolution of the system with moderate sizes. For larger 2D systems, we compare the mean occupation of the single-particle eigenstates for the ground and steady state computed through mean-field kinetic equations.
}

\vspace{10pt}
\noindent\rule{\textwidth}{1pt}
\tableofcontents\thispagestyle{fancy}
\noindent\rule{\textwidth}{1pt}
\vspace{10pt}

\section{Introduction}
\label{sec:intro}

A topological insulator is a state where fermionic particles completely fill a topologically non-trivial energy band~\cite{Hasan2010}. An important example is given by Chern insulators in two dimensions~\cite{Haldane, TKNN1982, Bernevig2013}, corresponding to lattice versions of the integer quantum Hall effect~\cite{vonKlitzing1980, PrangeGirvin1990}. Here particles occupy bands characterized by a non-zero Chern number, giving rise to quantized response functions, such as the Hall conductivity or a circular dichroism with respect to driving-induced interband excitations. 

In order to prepare a Chern insulator state in a quantum simulator of ultracold fermionic atoms in an optical lattice, two problems have to be addressed. On the one hand, topologically non-trivial Chern bands have to be engineered. This problem has been successfully addressed in a number of different experiments. Using Floquet engineering, both the paradigmatic square-lattice Harper-Hofstadter model as well as Haldane-type honeycomb lattice have been implemented~\cite{Jotzu2014, Aidelsburger2015, Tai2017, Wintersperger2020, Eckardt2017}. On the other hand, a band-insulating state has to be prepared, where (at least) one topological band has to be filled completely with fermions. This second problem is not yet solved fully. Namely, since (unlike electronic systems in solid state) ultracold atoms are not coupled to a thermal bath, the Chern insulator state has to be prepared adiabatically, starting from the topologically-trivial regime prepared initially. This implies that a topological phase transition has to be crossed, where the energy gap of the band structure closes. The band touching necessarily leads to a deviation from the desired adiabatic passage, corresponding to interband excitations that subsequently remain in the system due to its isolation. 

An appealing alternative to adiabatic passage is given by dissipative preparation methods~\cite{Diehl2008, Kraus2008, Verstraete2009}, in which a dissipative process is identified and engineered whose unique steady-state is the target state. These ideas have been theoretically explored, for instance, for preparing nonequilibrium Bose-Einstein condensates~\cite{Schnell2023, Petiziol2024c} and a range of topological states~\cite{Diehl2011,Goldman2016,Goldstein2019,Qin2018, PhysRevB.102.184302,Seetharam2015,Kapit2014,Liu_2021, Petiziol2022} including Floquet band insulators~\cite{Seetharam2015, Schnell2024} and fractional quantum Hall states \cite{Kapit2014,Liu_2021}.
In experiments, engineered dissipation has been employed to stabilize entangled states of ions~\cite{Barreiro2011},  
Mott insulating states of bosons~\cite{Tomitae1701513} and fermions~\cite{Sponselee_2018} in optical lattice
systems, and of photons in superconducting circuits~\cite{Ma2019}.
This approach has the twofold advantage that, on the one hand, the target state would be prepared independently of the initial state and that the system would return to the target state whenever driven away from it. 

In this work, we investigate the possibility to dissipatively prepare (topological) band insulators with ultracold atoms in optical lattices by means of Markovian feedback control~\cite{PhysRevA.47.642,Wiseman,wiseman_milburn_2009}. 
Measurement-based feedback control has already been successfully implemented in various systems for tasks such as quantum state preparation and stabilization~\cite{Doherty1999PRA,Wang2001PRA,Stockton2004PRA,Geremia2004Science,Geremia2006PRL,Yanagisawa2006PRL,Negretti2007PRL,Sayrin2011nature,Zhou2012PRL,Inoue2013PRL,Riste2013nature,Wade2015PRL,Cox2016PRL,Gajdacz2016PRL,Campagne2016PRL,Lammers2016PRA,Sudhir2017PRX,Fosel2018PRX,Borah2021PRL,Young2021PRR,Porotti2022quantum,Sivak2022PRX,Reuer2023NC,Hutin2025PRXQuantum},
cooling~\cite{PhysRevA.95.043641,Hush_2013,PhysRevA.80.013614,PhysRevA.75.051405,PhysRevA.100.063819,Ivanov_2014,PhysRevLett.96.043003,PhysRevLett.105.173003,PhysRevLett.111.103601,PhysRevLett.105.173003,Wilson2015,Yamaguchi2023PRA},
simulating nonlinear dynamics~\cite{PhysRevA.62.012307,PhysRevLett.124.110503,PhysRevA.102.022610},
controlling dynamics~\cite{PhysRevLett.110.013601,Vijay2012,PhysRevLett.88.093003,PhysRevLett.92.223004,PhysRevLett.110.210503,Kroeger_2020}
and manipulating phase transitions~\cite{Kopylov_2015,PhysRevLett.124.010603,PhysRevResearch.2.043325,PhysRevA.99.053612,Buonaiuto2021PRL}, among others.
As a measurement-based
approach, Markovian feedback control operates by continuously adding a signal-proportional feedback term to the system Hamiltonian~\cite{PhysRevA.47.642,Wiseman,wiseman_milburn_2009}. 
This method has been used for
  the stabilization of arbitrary one-qubit quantum states~\cite{Wang2001PRA,Campagne2016PRL}, the control of two-qubit entanglement~\cite{PhysRevA.71.042309,PhysRevA.76.010301,PhysRevA.78.012334,Wang2010}, and the generation of optical and spin squeezing~\cite{Thomsen2002PRA,Buonaiuto2021PRL}. In previous works by some of us, we have shown that Markovian
feedback control can be used to cool bosonic atoms in a one-dimensional optical lattice~\cite{lnwu}, and for quantum engineering of a synthetic thermal bath~\cite{2022arXiv220315670W} and heat-current-carrying states~\cite{Wu2022h}.
Here, we consider two-band fermionic models in one- (1D) and two-dimensional (2D) lattices at half filling. We first show that a mechanism able to dissipatively pump particles to the lower band is sufficient to prepare the desired state. Then, we discuss the experimental implementation of such an interband cooling process with measurement and homodyne-based feedback. We derive both an exact implementation and approximate ones that aim at favouring experimental feasibility. The approximate scheme further reveals interesting connections between the topological properties of the system, namely the unavailability of (exponentially) localized Wannier functions in topological bands, and the performance of the ground state preparation using only local measurements and feedback.

This paper is organized as follows.  A brief introduction to Markovian feedback control is given in Section~\ref{Markovian-feedback-control}. We then describe the basic idea of our cooling scheme in Section~\ref{basic-idea}, followed by the discussion of two approaches in Sections~\ref{sec:exact} and~\ref{sec:appr_meth} to construct the jump operator such that the  dissipative process drives the system towards the ground state. Our cooling scheme is benchmarked in Section~\ref{main-result}, where the two approaches are applied to different models, including the one-dimensional Rice-Mele model~[see Section~\ref{sec:1D}] and the two-dimensional Haldane model~[see Section~\ref{sec:2D}]. A summary of the main results is given in Section~\ref{conclusion} to conclude.

\section{Markovian Feedback control}
\label{Markovian-feedback-control}
Let us briefly recapitulate the idea of Markovian feedback control~\cite{PhysRevA.47.642,Wiseman,wiseman_milburn_2009}. Suppose we perform a continuous measurement of the observable $M$ on a system described by the Hamiltonian $H$. The system dynamics is then governed by the stochastic master equation~\cite{PhysRevA.47.642,Wiseman,wiseman_milburn_2009}~($\hbar=1$ hereafter),
\begin{equation}\label{SME}
	d\rho_c = -i[H,\rho_c] dt + {\cal D}[M]\rho_c dt + {\cal H}[M]\rho_c dW,
\end{equation}
where $\rho_c$ denotes the quantum state (density matrix) conditioned on the measurement result, with nonlinear superoperators
\begin{eqnarray}
{\cal D}[M]\rho &:=& M\rho M^\dag - \frac{1}{2}(M^\dag M \rho + \rho M^\dag M),\\
{\cal H}[M]\rho &:=& M\rho + \rho M^\dag - {\rm Tr}[(M+M^\dag)\rho]\rho\ ,
\end{eqnarray}
which describe the dissipation induced by the measurement, with $dW$ the standard Wiener increment with mean zero and variance $dt$. The measurement signal is given by
\begin{equation}
	I_{\rm hom} = {\rm Tr}[(M+M^\dag)\rho_c] + \xi(t),
\end{equation}
with $\xi(t)=dW/dt$.
By using the information obtained from the
measurements, one can introduce feedback control to the system such as to steer the system's dynamics for achieving desired effects. 

Here we consider the so-called Markovian feedback scheme introduced by Wiseman and Milburn~\cite{PhysRevA.47.642}, where the unprocessed measurement signal is fed back to the system by coupling it to an observable $F$. That is, we are introducing a term $I_\text{\rm hom}F$ to the system.
 For such a feedback, the delay time between
the measurement and the application of the control field is assumed to be negligible compared to the typical
timescales of the system. 
For instance, in cold atom experiments, the typical time scales~(such as tunneling time) are on the order of milliseconds~\cite{Bloch2012}. Hence, a control on the higher kHz scale is sufficient, which can be achieved easily by using digital signal processors. 

According to Markovian feedback control theory~\cite{Wiseman,wiseman_milburn_2009},
the combined action of measurement and feedback results in an effective dissipative process described by the feedback-modified stochastic master equation
\begin{eqnarray}\label{sme_fbM}
	d\rho_c &=& -i[H+H_{\rm fb},\rho_c]dt + {\cal D}[C]\rho_c dt +{\cal H}[C]\rho_c dW,
\end{eqnarray}
with the quantum jump operator
\begin{equation}\label{A}
	C = M-iF,
\end{equation}
and feedback-induced term $H_{\rm fb}=\frac{1}{2}(M^\dag F+F M)$.
By taking the ensemble average of the possible measurement outcomes, we arrive at the Wiseman-Milburn master equation
\begin{equation}\label{me_fbM0}
	\frac{d\rho}{dt} = -i[H + H_{\rm fb},\rho] + {\cal D}[C]\rho.
\end{equation}
Suppose the measurement strength is $\gamma$, i.e., $M \propto \sqrt{\gamma}$. For the feedback, we assume $F \propto \sqrt{\gamma}$ so that the jump operator $C \propto \sqrt{\gamma}$, and the feedback-induced term $H_{\rm fb}$ is on the order of $\gamma$. In this work, we consider weak measurement, with $\gamma$ small enough such that $H_\mathrm{fb}$ has negligible impact as compared to the system Hamiltonian $H$. 
Excluding the impact of $H_\mathrm{fb}$, the steady state of Eq.~\eqref{me_fbM0} is given by
\begin{equation}\label{me_fbM}
	{\cal L}\rho \equiv -i[H,\rho] + {\cal D}[C]\rho = 0.
\end{equation}
For weak measurement, the steady state is well approximated by a mixture of eigenstates of the system, with the weight dependent on the specific form of the jump operator $C$. In the following, we will discuss how to design $C$ to achieve a target steady state.

\section{Constructing measurement and feedback operators}

\subsection{Basic idea}\label{basic-idea}
 We consider discrete tight-binding models for fermionic ultracold atoms in an optical lattice at half filling. We focus on systems whose single-particle energy spectrum is characterized by two energy bands with non-trivial topological properties, such that the ground state $\ket{g}$ at half filling is a topological insulator.
Our goal is to design the jump operator $C$, i.e., the underlying measurement and feedback, such that the effective dissipative dynamics drives the system towards its many-body ground state $\ket{g}$. The system~\eqref{me_fbM} has a pure steady state if the effective Hamiltonian $H_{\rm eff} = H -iC^\dag C/2$ and the collapse operator $C$ have a common eigenstate~\cite{PhysRevA.72.024104}, which is then the steady state. Hence, the ground state $|g\rangle$ will be a steady state of the system if it is a dark state of the jump operator $C$, i.e., 
\begin{equation}\label{cooling_condition}
    C\ket{g}=0.
\end{equation}
Therefore, we need to construct a jump operator which satisfies the condition~\eqref{cooling_condition}.
It is intuitive that such a dissipation process will drive the system towards the ground state, since the coupling between the ground state and other eigenstates is unidirectional, i.e., the transfer from other eigenstates to the ground state is allowed, while the reverse channel is blocked, as indicated by~\eqref{cooling_condition}. 

For the cooling scheme to be successful, the ground state should be the {\emph {unique}} steady state of the system.
Although this property in general depends on the details of the model considered, a general feature that can yield multiple steady states, spoiling the state preparation protocol, is degeneracy. For instance, consider two degenerate many-body eigenstates $\ket{\psi}$ and $\ket{\psi'}$ that are mapped by the jump operator to the same state $\ket{\tilde{\psi}} = C\ket{\psi}=C\ket{\psi'}$. Then, the superposition $\ket{\psi_-} = (\ket{\psi}-\ket{\psi'})/\sqrt{2}$ will be both an eigenstate of $H_{\rm eff}$ and of the jump operator with eigenvalue 0, $C\ket{\psi_-}=0$. Therefore, $|\psi_-\rangle$ will also be a dark state like $|g\rangle$. Being a common eigenstate of $H_{\rm eff}$ and $C$, it thus constitutes a steady state of the system. A key point for the success of our cooling scheme is, therefore, to ensure that there is no degeneracy in the system. In order to get rid of degeneracies in the example models, we will employ different strategies detailed in \ref{section:lift_degeneracy}.

In the following, we present two different choices for constructing a suitable jump operator $C$. We first discuss an \textit{exact} construction, which is however challenging to implement experimentally. Starting from this, we then derive a second \textit{approximate} construction for the purpose of enhancing the experimental feasibility. 

\subsection{Exact construction} \label{sec:exact}

We start by observing that, since the target state features all particles occupying the lower band only and filling all single-particle states therein, the jump operator $C$ must be able to deplete particles from the upper band and pump them to the lower band. It is further needed that particles can be pumped to any state in the lower band: this is guaranteed if the dissipative process provides non-zero transition matrix elements between any state in the upper band to any state in the lower band. Finally, we note that, while particles are pumped to the lower band, Pauli's exclusion principle will take care of inducing a uniform distribution of particles throughout the lower band, until the target state is eventually reached, by forbidding multiple occupancy. 
A jump operator complying with these conditions can be constructed as 
\begin{equation} \label{eq:jump_exact}
C = \sqrt{\gamma} C_{-}^\dagger C_+, 
\end{equation}
where $C_{\pm}$ destroys ($C_{\pm}^\dagger$, creates) a particle in states $\ket{\pm}=C_{\pm}^\dagger\ket{0}$ that have overlap with all states in the upper (+) and in the lower ($-$) band, respectively. In this way, the jump process transfers a particle from the upper band to the lower band.
Concretely, we will choose $\ket{\pm}$ to be Wannier-like states for each band. Given the Bloch states $\ket{k_{\pm}}$ of the upper and lower band in a system with $N$ unit cells, the Wannier-like states are defined as
\begin{equation}
\ket{\pm} = \frac{1}{\sqrt{N}} \sum_{k\in\mathcal{B}} e^{i\varphi_{k_{\pm}}} \ket{k_{\pm}},
\end{equation}
where $\mathcal{B}$ is the first Brillouin zone and $e^{i\varphi_{k_{\pm}}}$ is a gauge factor associated to the Bloch state $\ket{k_{\pm}}$. By appropriate choice of $e^{i\varphi_{k_{\pm}}}$ we can obtain well-localized Wannier states. In terms of creation and annihilation operators for Bloch states $c_{k,\pm}$, such that $\ket{k_{\pm}}=c_{k,\pm}^\dagger \ket{0}$, the jump operator then reads 
\begin{equation} \label{eq:Abloch}
C = \frac{\sqrt{\gamma}}{N} \sum_{k, k' \in \cal{B}} e^{i(\varphi_{k'_{+}}-\varphi_{k_{-}})} c_{k,-}^\dagger c_{k',+},
\end{equation}
with $\gamma$ the measurement strength. It does indeed have a matrix element connecting any state in the upper band to any states in the lower band, as desired. 

Within the Markovian feedback formalism, the collapse operator $C$ is related to the measurement operator $M$ and feedback operator $F$ via $C = M - iF$ in Eq.~(\ref{A}), where both $M$ and $F$ are required to be Hermitian. Combining this with its Hermitian conjugate $C^\dagger = M + iF$, we obtain the explicit expressions: $M = (C+C^\dagger)/2$ and $F=i(C-C^\dagger)/2$. From the jump operator $C$ given in Eq.~\eqref{eq:Abloch}, the corresponding measurement and feedback operators can be identified as
\begin{align} \label{eq:M_operator}
M = & \frac{\sqrt{\gamma}}{2N}\sum_{k,k'\in\mathcal{B}} \Big(e^{i(\varphi_{k'_{+}}-\varphi_{k_{-}})}c_{k,-}^\dagger c_{k',+} + e^{i(-\varphi_{k'_{+}}+\varphi_{k_{-}})}c_{k,-} c_{k',+}^\dagger \Big) \ , \\
\label{eq:F_operator}
F = & \frac{i\sqrt{\gamma}}{2N}\sum_{k,k'\in\mathcal{B}} \Big( e^{i(\varphi_{k'_{+}}-\varphi_{k_{-}})}c_{k,-}^\dagger c_{k',+} -  e^{i(-\varphi_{k'_{+}}+\varphi_{k_{-}})}c_{k,-} c_{k',+}^\dagger \Big) \ .
\end{align}
Even though the Wannier states can be chosen to be localized in real space (exponentially for topologically trivial bands and like a power-law  for topologically non-trivial ones), they still span over the whole lattice. 
This may constitute a challenge for the experimental implementation of this approach, and motivates the search for a less demanding strategy, which we develop in the next section.
For the exact construction of collapse operator, we set the gauge factor to be 0, since the localization of the Wannier states should not influence our cooling scheme in this case.

\subsection{Approximate construction} \label{sec:appr_meth}
In experiments, the measurement of on-site population in Eq.~(\ref{eq:M_operator}) can be implemented via homodyne detection of
the off-resonant scattering of structured probe light from the atoms \cite{PhysRevLett.114.113604,PhysRevLett.115.095301,RevModPhys.85.553} and the feedback operator in Eq.~(\ref{eq:F_operator}) with a complex tunneling can be realized by accelerating the lattice~\cite{RevModPhys.89.011004,lnwu, Jotzu2014,Wintersperger2020}. In the \emph{exact} scheme, the feedback operator $F$ is typically \emph{non-local} in real space, reflecting the delocalized nature of the optimal jump operator $C$. Implementing such non-local operations directly is a significant experimental challenge with current technologies, as it would require coherent control over spatially separated sites. While one could envision engineered feedback using global fields shaped by spatial light modulators or programmable optical potentials, these approaches are still limited in resolution and scalability. Thus, the exact scheme serves primarily as a theoretical benchmark that demonstrates the best possible performance under idealized feedback. 

 In order to favour experimental realizations, we develop a second approach in which the measurement and feedback operators are strictly constrained to have support on only a few neighbouring lattice sites. The idea is based on the fact that, by adjusting the gauge factors $e^{i\varphi_{k_{\pm}}}$ in Eq.~\eqref{eq:Abloch}, the Wannier states can be chosen to be well localized in space, see more details in \ref{section:MLWF}. This approximate scheme is designed to preserve good performance while remaining implementable with current or near-term capabilities. In particular, the feedback operator $F$ will typically correspond, in this case, to local tunneling Hamiltonians acting within a small spatial region (e.g., a unit cell), with complex amplitudes, which could be implemented experimentally using techniques such as Floquet engineering~\cite{Jotzu2014,Wintersperger2020}. We start by considering a jump operator $C_\ell$ entirely localized in the $\ell$-th unit cell. We further use labels $A$ and $B$ for the two inequivalent sites in the unit cell, and $a_{\ell,A}$ ($a_{\ell,B}$) annihilates particles at site $A$ ($B$) in the $\ell$-th unit cell. We  then choose an {\it ansatz} for the constrained jump operator of the form 
 \begin{equation}\label{Cl}
C_\ell = \sqrt{\gamma} b_{\ell,-}^\dagger b_{\ell,+},
 \end{equation}
 with operators
\begin{equation} \label{eq:jump_approx}
b_{\ell,+} = \cos(\xi) a_{\ell,A} + \sin(\xi) a_{\ell,B}, \qquad b_{\ell,-} = -\sin(\xi) a_{\ell,A} + \cos(\xi) a_{\ell,B} \ .
\end{equation}
These operators annihilate particles in states
$\ket{b_{\ell,\pm}}=b_{\ell,\pm}^\dagger\ket{0}$, i.e.,
 \begin{equation}\label{ansatz}
  \ket{b_{\ell,+}} = \cos(\xi) |{\ell,A}\rangle + \sin(\xi) |{\ell,B}\rangle, \quad  \ket{b_{\ell,-}} = -\sin(\xi) |{\ell,A}\rangle + \cos(\xi) |{\ell,B}\rangle,
 \end{equation}
respectively, that are generic real superpositions of the states localized at $A$ and $B$ site parametrized in terms of an angle $\xi$. We choose real superpositions, such that they involve only one free paratemer $\xi$. For the operator $C_\ell$ to successfully pump particles from the upper band to the lower band,  we aim at selecting superposition states $\ket{b_{\ell,\pm}}$ such that $\ket{b_{\ell,+}}$ has maximal (minimal) overlap with the upper (lower) band, while $\ket{b_{\ell,-}}$ has maximal (minimal) overlap with the lower (upper) band. To find such states (the optimal angle $\xi$) we resort to analytical and numerical optimization schemes.

In the applications treated in the following, we will also consider jump operators constrained to a larger number of sites. Given a set $\mathcal{S}_n$ of $n$ sites to which the jump operator $C_{\mathcal{S}_n}= (b_-^{\mathcal{S}_n})^\dagger b_+^{\mathcal{S}_n}$ is constrained, we choose the {\it ansatz}
\begin{equation} \label{eq:ansatz_WF_nsites}
b_\pm^{\mathcal{S}_n} = \sum_{j\in\mathcal{S}_n} \beta_{\pm,j} a_{j},
\end{equation}
where the index $j$ indicates real-space coordinates. 
For the one-dimensional systems considered in the following, we can find optimal $\ket{b_{\ell,\pm}}$ analytically for $n=2$. For larger $n$ and in other applications, we will find optimal coefficients $\beta_{\pm,j}$ through numerical optimization, minimizing the overlap of $\ket{b_-^{\mathcal{S}_n}}$ ($\ket{b_+^{\mathcal{S}_n}}$) with the upper (lower) band. 

\section{Application to paradigmatic models}\label{main-result}

We will now characterize the performance of the exact and approximate approaches for the preparation of the ground state for two different classes of models. We start from one-dimensional models, the Su-Schrieffer-Heeger~\cite{PhysRevLett.42.1698} and the Rice-Mele model~\cite{RiceMele}, respectively, see Section~\ref{sec:1D}. Their one-dimensional band-structures (at fixed parameters) are not characterized by Chern numbers. However, they provide a minimal proof-of-principle scenario that allows us to test the scheme proposed above. In a second step, we investigate the two-dimensional Haldane model~\cite{Haldane} both in its topologically trivial and non-trivial regime, see Section~\ref{sec:2D}. 
 
For the aforementioned models of moderate size, the steady state of the system is obtained by numerically solving Eq.~\eqref{me_fbM} through exact diagonalization or via time evolution for a time sufficiently long to ensure that the system has reached a steady state [Section~\ref{small-system}]. Note that the term $H_{\rm fb}$ is neglected throughout the paper, as justified in  Appendix~\ref{section:change_gamma}.
To quantify the performance of our cooling scheme, we calculate the fidelity between the steady state $\rho_{\rm ss}$ and the ground state $|g\rangle$ of the system, which is defined as
\begin{equation}\label{fidelity}
    \mathcal{F} = \sqrt{\bra{g}\rho_{\rm ss}\ket{g}}\ ,
\end{equation}
and takes values $0\le \mathcal{F}\le 1$. A larger fidelity implies a better performance of our scheme.
 
Furthermore, in order to treat larger systems, we derive kinetic equations of motion using a mean-field approach for the case of the Haldane model [Section~\ref{mean-field}].

\subsection{One-dimensional systems: The Rice-Mele and Su-Schrieffer-Heeger models} \label{sec:1D}

\begin{figure}[t]
	\includegraphics[trim = 28mm 97mm 32mm 43mm, clip, width=1
    \textwidth]{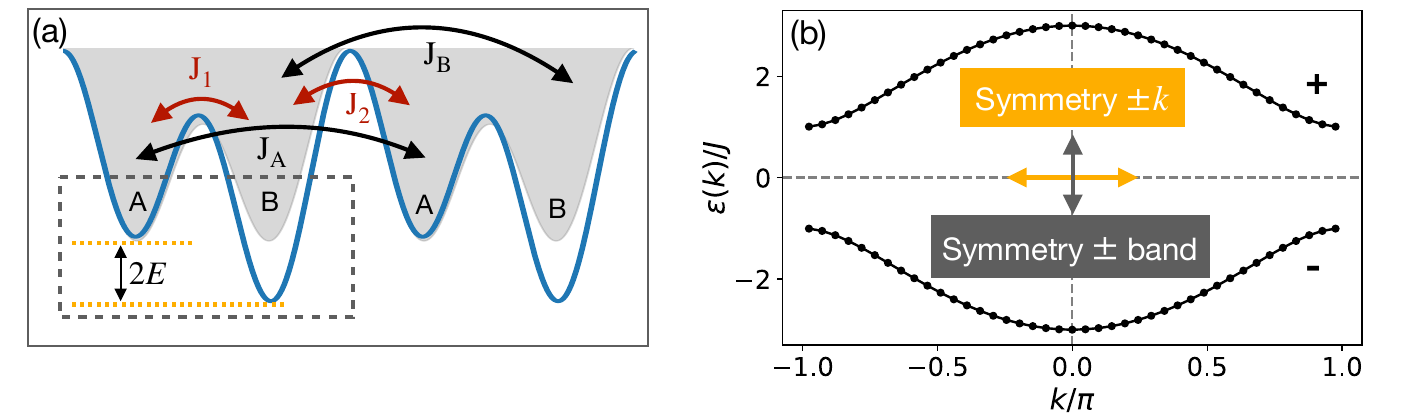}
	\caption{(a) A sketch of the Rice-Mele model in blue color, characterized by a staggered onsite potential, intracell hopping amplitude $J_1$, and intercell
hopping amplitude $J_2$.  
$J_{A}$ and $J_{B}$ denote the next nearest neighbour hopping amplitudes.
The potential can be realized by a superlattice, Eq.~\eqref{eq:lattice_potential_rice_mele}, in the experiments.  The black dashed line indicates a unit cell. For $E=0$ one obtains the Su-Schrieffer-Heeger model as a filled gray color, which is characterized by a symmetric double-well system. (b) Single-particle energy spectrum for the SSH model with periodic boundary condition, i.e., Eq.~\eqref{ssh-spectrum} for $E=0$ and $J_1=2J_2=2J$. The symmetries related to $\pm k$ and between upper(+)/lower (-) bands are shown.}
	\label{fig:rice_mele_ssh_potential}
\end{figure}

Our first example is the Rice-Mele model \cite{RiceMele,Asb_th_2016}, which describes a one-dimensional~(1D) tight-binding chain with staggered hopping parameters and staggered on-site potentials [see sketch in Fig. \ref{fig:rice_mele_ssh_potential}(a)]. The Hamiltonian reads 
\begin{equation}
	\label{eq:rice_mele_Hamiltonian}
	H = \sum_{l}\left[ E(n_{l,A} - n_{l,B}) -\big(J_1 a_{l,A}^\dagger a_{l,B}+J_2 a_{l,B}^\dagger a_{l+1,A}+ \mathrm{h.c.}\big) \right],
\end{equation}
with $n_{l,\nu}= a_{l,\nu}^\dagger a_{l,\nu}$ being number operators for $\nu=A,B$. The chain consists of $N$ unit cells, each unit cell hosting two sites: one on sublattice $A$, one on sublattice $B$. Here, $J_1$ and $J_2$ denote the intracell and intercell nearest-neighbour tunnelling strength, and $E$ is the staggered onsite potential. This model can be realized with bichromatic ultracold atoms in optical lattices~\cite{Nakajima_2016}. By using two counter propagating laser beams with period $d$ and $d/2$ respectively, one can obtain a superlattice with potential 
\begin{equation}
	\label{eq:lattice_potential_rice_mele}
	V(x)=V_1 \sin^2\left(\pi \frac{x}{d}\right)+V_2 \sin^2\left(2\pi \frac{x}{d} + \theta\right),
\end{equation} 
where 
$\theta$ is the phase difference between these two different sublattices, and $V_1$ and $V_2$ are the lattice depths for the wider and narrower lattice, respectively. 

For a translationally invariant chain with periodic boundary conditions~(PBC), transforming the Hamiltonian to momentum space via $a_{k, A/B}^\dagger=\frac{1}{\sqrt{N}}\sum_\ell a_{\ell,A/B}^\dagger e^{i k \ell}$, where $k =q \frac{2\pi}{N}$ with $q = -\frac{N-1}{2},..., \frac{N-1}{2}$, yields 
\begin{equation}
H = \sum_{k \in \mathcal{B}} (a_{k,A}^\dagger, a_{k,B}^\dagger)\mathfrak{h}(k) \begin{pmatrix}
a_{k, A}\\ a_{k,B}
\end{pmatrix},
\end{equation}
characterized by the single-particle Hamiltonian
\begin{equation}
	\label{eq:ssh_momentumspace_hamiltonian}
	\mathfrak{h}(k)= \begin{pmatrix}
		E & -J_1-J_2 e^{ik} \\
		-J_1-J_2 e^{-ik} & -E
	\end{pmatrix},
\end{equation}
whose eigenenergies read 
\begin{equation}\label{ssh-spectrum}
\varepsilon(k) = \pm \sqrt{E^2 + (J_1+J_2\cos  k)^2+(J_2\sin k)^2},
\end{equation}
with the spectrum shown in Fig.~\ref{fig:rice_mele_ssh_potential}(b) for $E=0$ and $J_1=2J_2$. 
The single-particle spectrum~\eqref{ssh-spectrum} has a symmetry with respect to $\pm k$
and a symmetry between upper and lower band. The 
former results from time-reversal invariance and the 
latter is a consequence of
chiral symmetry, i.e., $\sigma_z {\frak h}(k)\sigma_z=-{\frak h}(k)$, with $\sigma_z$ being the Pauli matrix. Due to these symmetries, the many-body spectrum will be degenerate. For an effective cooling scheme, we need to lift the degeneracies in the half filling spectrum, which could be realized in different ways, such as by introducing a weak nearest-neighbour (NN) interaction under open boundary condition (OBC), or by introducing next-nearest-neighbour (NNN) tunneling under OBC, see more in \ref{section:lift_degeneracy}.

\subsubsection{Feedback cooling of the SSH chain}\label{cooling_ssh}

For $E=0$ in the Hamiltonian \eqref{eq:rice_mele_Hamiltonian}, the Rice-Mele model reduces to the Su-Schrieffer-Heeger (SSH) model~\cite{PhysRevLett.42.1698, Asb_th_2016}. The Su-Schrieffer-Heeger (SSH) model shows a topological phase transition at $J_1/J_2 = 1$, separating a topologically trivial phase  for $J_1/J_2 >1$ from a non-trivial one for $J_1/J_2 <1$, which is characterized by a non-trivial winding of the Berry phase through the Brillouin zone. To study the effectiveness of the exact method, we first consider open SSH chains of moderate size ($N=1 \sim 4$) at half filling that allow us to numerically solve the steady-state equation $\mathcal{L}\rho =0$, with the superoperator $\mathcal{L}$ given in Eq.~\eqref{me_fbM} and jump operator $C$ of Eq.~\eqref{eq:jump_exact}. To lift degeneracies in the many-body spectrum (see Section \ref{basic-idea}), we further introduce a weak nearest-neighbour interaction
\begin{equation} \label{eq:Hint}
\hat{H}_I=U\sum_\ell( n_{\ell-1,B} n_{\ell,A}+ n_{\ell,A} n_{\ell+1,B}).
\end{equation}
We solve Eq.~\eqref{me_fbM} numerically using the toolbox QuTiP~\cite{Johansson_2013} in Python to find the steady state. As expected by construction, the exact method always yields the desired state $\ket{g}$ with close-to-unity fidelity between the steady state and the ground state for all system sizes investigated numerically, both in the topological~($J_1/J_2<1$) and trivial~($J_1/J_2>1$) phase.

Considering now the approximate method, we construct analytically optimal states $\ket{b_{\ell,\pm}}$ of Eq.~\eqref{eq:jump_approx} which are used to define the jump operator $C_\ell$ of Eq.~\eqref{Cl} localized in the $\ell$-th unit cell. In particular, we search for states such that $\ket{b_{\ell,-}}$ ($\ket{b_{\ell,+}}$) has minimal overlap with the upper (lower) band. The eigenvectors of the single-particle momentum-space Hamiltonian \eqref{eq:ssh_momentumspace_hamiltonian} for $E=0$ can be written as follows,
\begin{equation} \label{eq:eigenVector_ssh}
\ket{k_\pm} = \frac{1}{\sqrt{2}}\begin{pmatrix} \pm e^{i \phi_k} \\ 1 \end{pmatrix}, \qquad  e^{i \phi_k } = \frac{-J_1-J_2 e^{ik}}{|-J_1-J_2 e^{ik}|} . \end{equation}
The squared overlap of state $\ket{k_\pm}$ with $\ket{b_{\ell,-}}$ is thus
\begin{equation}\label{overlap_k}
|\!\braket{k_\pm\vert b_{\ell,-}}\!|^2 = \frac{1}{2N}\big[1 \mp \sin(2\xi)\cos(\phi_k)\big].
\end{equation}
Hence, the total overlap of the state $\ket{b_{\ell,-}}$  with the upper ($+$) and lower ($-$) band, shall be defined by the probability
\begin{align} \label{eq:boverlap}
\mathcal{P}_{\pm}(\xi) = \sum_{k \in \mathcal{B}} |\!\braket{k_{\pm}\vert b_{\ell,-}}\!|^2 & = \frac{1}{2}\mp \frac{\sin(2\xi)}{2N}\sum_{k\in\mathcal{B}} \cos(\phi_k),\\
& =\frac{1}{2}\mp \frac{\sin(2\xi)}{2N}s. \label{eq:secline}
\end{align}
For convenience, in Eq.~\eqref{eq:secline} we defined $s \equiv \sum_{k\in\mathcal{B}}\cos(\phi_k)$, which is always negative for a large system with $J_1 \neq 0$ and vanishes for $J_1=0$, since the $k$ values are equally spread over the first Brillouin zone. 

Considering first $N=1$ for simplicity, we have $s=-1$. In this case, the minimal~(maximal) overlap of $\ket{b_{\ell,-}}$ with the upper~(lower) band is attained at $\xi=-\pi/4$, leading to states 
\begin{equation}\label{blm}
  \ket{b_{\ell,\pm}} = \frac{1}{\sqrt{2}}( \ket{\ell,A} \mp \ket{\ell,B}). 
\end{equation}
The orthogonality between $\ket{k_+}$ and $\ket{k_-}$ and between $\ket{b_{\ell,+}}$ and $\ket{b_{\ell,-}}$ then guarantees that $|\!\braket{k_+|b_{\ell,-}}\!|^2=|\!\braket{k_-|b_{\ell,+}}\!|^2$, such that $\ket{b_{\ell,+}}$ has minimal overlap with the lower band. We discuss below that this construction works well numerically also for larger $N$ for $J_1/J_2\gg 1$, while the regime $J_1/J_2\ll 1$ can be treated as well by a shift of $\ket{b_{\ell,\pm}}$ by one lattice site.
With this choice for the states $\ket{b_{\ell,\pm}}$, the jump operator $C_\ell$ of Eq.~\eqref{Cl} is determined as
\begin{align} \label{eq:collapse_two_sites}
    C_\ell = \frac{\sqrt{\gamma}}{2} (n_{\ell,A}-n_{\ell,B}) + \frac{\sqrt{\gamma}}{2}(a_{\ell,B}^\dag a_{\ell,A} - a_{\ell,A}^\dag a_{\ell,B}),
\end{align}
The corresponding measurement and feedback operators thus read
\begin{equation} 
M_\ell= \frac{\sqrt{\gamma}}{2} (n_{\ell,A}-n_{\ell,B}), \qquad F_\ell= -i\frac{\sqrt{\gamma}}{2}(a_{\ell,A}^\dag a_{\ell,B} - a_{\ell,B}^\dag a_{\ell,A}).
\end{equation}
Here, $M_\ell$ measures the population imbalance between the two sites $A$ and $B$ in the $\ell$-th unit cell, and $F_\ell$ denotes a tunneling between these two sites with a complex amplitude.
The measurement of on-site population can be implemented via homodyne detection of
the off-resonant scattering of structured probe light from the atoms~\cite{PhysRevLett.114.113604,PhysRevLett.115.095301,RevModPhys.85.553}.
The tunneling with a complex amplitude can be realized by accelerating the lattice~\cite{RevModPhys.89.011004,lnwu}.

\begin{figure}[t]
		\includegraphics[trim = 0mm 82mm 20mm 46mm, clip, width=1
    \textwidth]{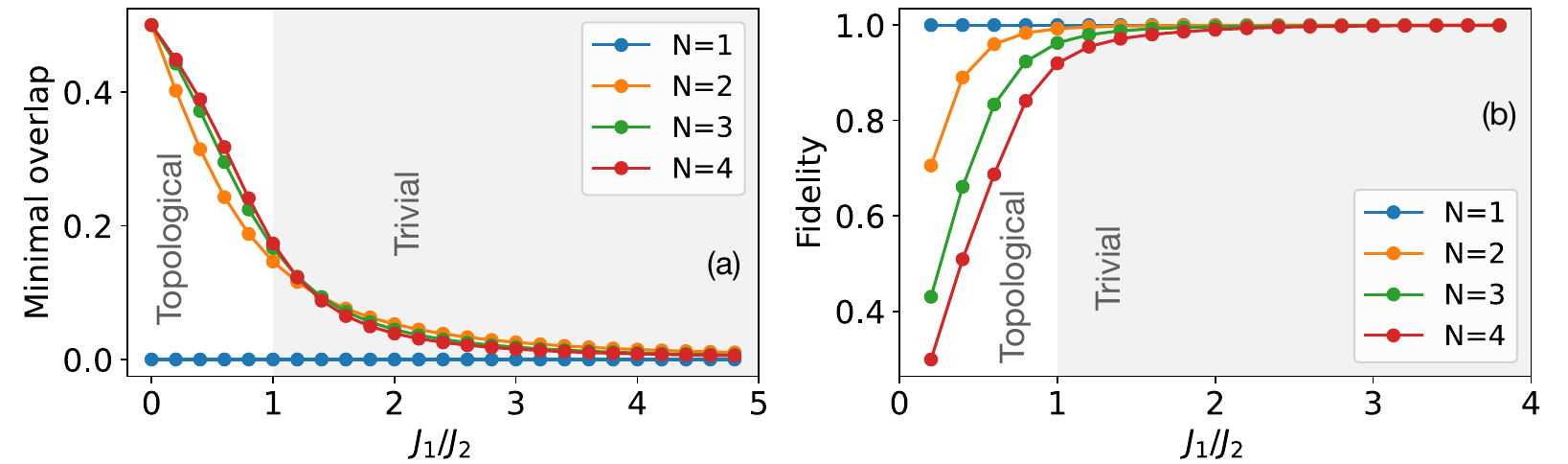}
		\caption{The approximate approach for the SSH model. The gray shaded background color indicates the trivial phase and the white background color indicates the topological phase. (a) The minimal overlap of Eq.~\eqref{eq:boverlap} between the state 
		$\ket{b_{\ell,-}}$ of lower band given by Eq.~\eqref{blm}
		and the upper band $\ket{k_{+}}$ for different small systems with $N$ unit cells as a function of $J_1/J_2$. (b) Fidelity defined in Eq.~\eqref{fidelity} as a function of $J_1/J_2$ with open boundary conditions for different small systems with $N$ unit cells.
		The steady state $\rho_{\rm ss}$ is obtained by solving Eq.~\eqref{me_fbM} with the jump operator constrained to two sites within one unit cell given by Eq.~\eqref{eq:collapse_two_sites}. The measurement strength is set to $\gamma=0.0001J$ with $J$ the energy unit.
		}%
		\label{fig:sshapproachII_fidelitydifferentunitcellnumbers}%
\end{figure}

Figure~\ref{fig:sshapproachII_fidelitydifferentunitcellnumbers}(a) shows the minimal overlap $\sum_k |\!\braket{k_+|b_{\ell,-}}\!|^2$ between the state $|b_{\ell, -}\rangle$ given in Eq.~\eqref{blm} and the whole upper band. Figure \ref{fig:sshapproachII_fidelitydifferentunitcellnumbers}(b) shows instead the fidelity given by the approximate cooling protocol as a function of $J_1/J_2$ for different system sizes ranging from 1 to 4 unit cells. As long as $J_1/J_2>1$ the fidelity is close to unity, while this is not the case for $J_1/J_2<1$ instead.  
This can easily be understood by considering that for $J_1 > J_2$ the system dimerizes, with the dimers localized in each unit cell. Maximally localized Wannier functions of the form of $\ket{b_{\ell,\pm}}$ [Eq.\ \eqref{blm}] can thus be constructed, which are mainly localized within the $\ell$th unit cell (\ref{section:MLWF}). For $J_1 < J_2$, instead, the dimers straddle two neighbouring unit cells \cite{Asb_th_2016} and, in this case, the maximally localized Wannier functions also straddle two unit cells. Following this reasoning, successful cooling in the topological phase $J_1/J_2<1$ can also be achieved by choosing a collapse operator localized on neighbouring sites belonging to different unit cells. This can also be seen formally from Eq. \eqref{eq:boverlap}: in the limit $J_2=0$, it holds that $\cos(\phi_k)=-1$ for any value of $k$, so the probability overlap ${\cal P}_\pm(\xi)=[1 \pm \sin(2\xi)]/2$ can reach $0$ or $1$ for an appropriate choice of $\xi$, while in the limit $J_1=0$ it becomes $1/2$ for $N>1$ and is independent from $\xi$.

\subsubsection{Rice-Mele pumping cycle}

We now construct the approximate jump operator $C_\ell$ for the Rice-Mele model of Eq.~\eqref{eq:rice_mele_Hamiltonian} with a non-zero staggered on-site potential $E\ne 0$. A state in the upper band can be written in the form
\begin{equation}
\ket{k_+} = \cos\frac{\chi_k}{2}\ket{k,A} + e^{-i\phi_k}\sin\frac{\chi_k}{2}\ket{k,B},
\end{equation}
where $\chi_k = 2\arctan \big[(\sqrt{E^2+J^2}-E)/J \big]$ with $-J_1-J_2 e^{ik}=Je^{i\phi_k}$. The squared overlap $\mathcal{P}_+(\xi)$ of $\ket{b_{\ell,-}}$ defined in Eq.~\eqref{eq:jump_approx} with the whole upper band is then
\begin{equation}
\mathcal{P}_+(\xi) = \frac{1}{2}\left\{1-  \frac{1}{N}\cos(2\xi)\sum_{k\in\mathcal{B}}\cos\chi_k-\frac{1}{N}\sin(2\xi) \sum_{k\in \mathcal{B}}\sin\chi_k \cos(\phi_k) \right\} .
\end{equation}
If $\sum_{k\in\mathcal{B}}\cos\chi_k\ne 0$ and $2\xi\ne (n+1/2) \pi$ with integer $n$, the extremal point satisfies
\begin{equation}\label{eq.xi}
\tan{2\xi}= \frac{\sum_{k\in \mathcal{B}}\sin\chi_k \cos(\phi_k)}{\sum_{k\in \mathcal{B}}\cos\chi_k}.
\end{equation}
With the approximate cooling operator determined by $\xi$, we study the cooling during a Rice-Mele pumping cycle. In this process, the parameters $E$, $J_1$ and $J_2$ in the Hamiltonian~\eqref{eq:rice_mele_Hamiltonian} are modulated periodically in time in a slow, adiabatic fashion. Through an appropriate parameter variation, the modulation pumps an integer number of particles along the chain which is determined by the Chern number of the valence band~\cite{Thouless1983}. This so-called topological charge pump that can be thought of as a 1D analog of the quantum Hall effect, where one spatial dimension is substituted by the temporal one~\cite{Kitagawa2010}.
\begin{figure}[t]
		\includegraphics[trim = 0mm 94mm 0mm 35mm, clip, width=1
    \textwidth]{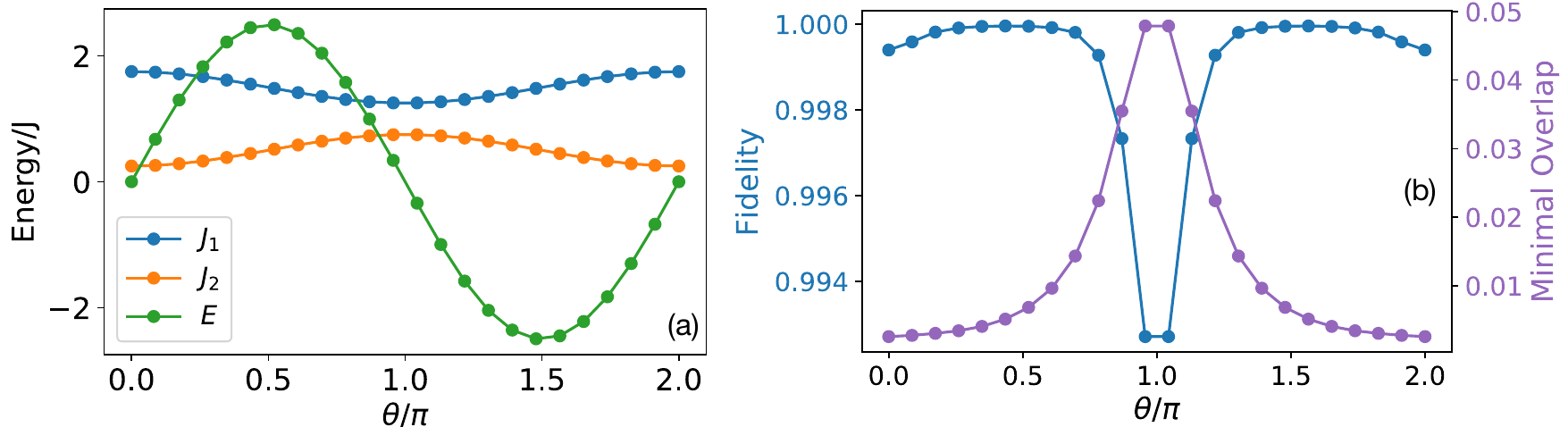}
		\caption{The approximate method for a Rice-Mele pumping cycle. (a) Hopping amplitudes $J_1$, $J_2$ and onsite potential $E$ as a function of $\theta$ given by Eq.~\eqref{RM-parameter}. (b) The minimal overlap and the fidelity defined in Eq.~\eqref{fidelity} during the pumping cycle as a function of $\theta$ for a small system with 3 unit cells under open boundary condition. The steady state $\rho_{\rm ss}$ is given by Eq.~\eqref{me_fbM} and the jump operator constrained to two sites within one unit cell is given by Eq.~\eqref{eq:collapse_two_sites}. The measurement strength is set to $\gamma=0.0001J$ with $J$ the energy unit. The steady state is obtained by solving Eq.~\eqref{me_fbM}.} 
	   \label{fig:ricemeleapproach2pumpingcycle}
\end{figure}
We consider the following modulation of the on-site potential and hopping parameters~\cite{Nakajima_2016,Nakajima_2021},
\begin{align}\label{RM-parameter}
&E(\theta) = \frac{5J}{2}\sin\theta \ , \qquad J_1(\theta) = \frac{J}{2}\left[ 3 + \frac{1}{2}\cos\theta \right]\ ,\qquad J_2(\theta) = \frac{J}{2}\left( 1 - \frac{1}{2}\cos\theta\right)\
\end{align}
as depicted in Fig.~\ref{fig:ricemeleapproach2pumpingcycle}(a), with $J$ the energy unit. In an experiment, the modulation is achieved by varying the phase difference $\theta$ between the two sublattices, see Eq. \eqref{eq:lattice_potential_rice_mele}.
The cooling fidelity is reported in Fig.~\ref{fig:ricemeleapproach2pumpingcycle}(b) as a function of $\theta$. In a large parameter regime, the deviation from unity is less than $10^{-3}$ and it remains always well below one percent, indicating that the approximate cooling protocol is reliable during the whole pumping cycle. The fidelity is slightly lower (although by a few parts in a thousand only) for values around $\theta\simeq \pi$. This effect can be explained by noting that at such values the on-site potential is close to zero. Non-zero values of $E$ indeed contribute in effectively reducing the hopping amplitude by bringing nearby sites off-resonant, thus favouring localized states that are well described by the single-cell ansatz~\eqref{ansatz} used for $\ket{b_{\ell,\pm}}$.
\subsection{Two-dimensional topological insulator: The Haldane model} \label{sec:2D}
\begin{figure}[t]
	\includegraphics[trim = 25mm 78mm 17mm 38mm, clip, width=1
    \textwidth]{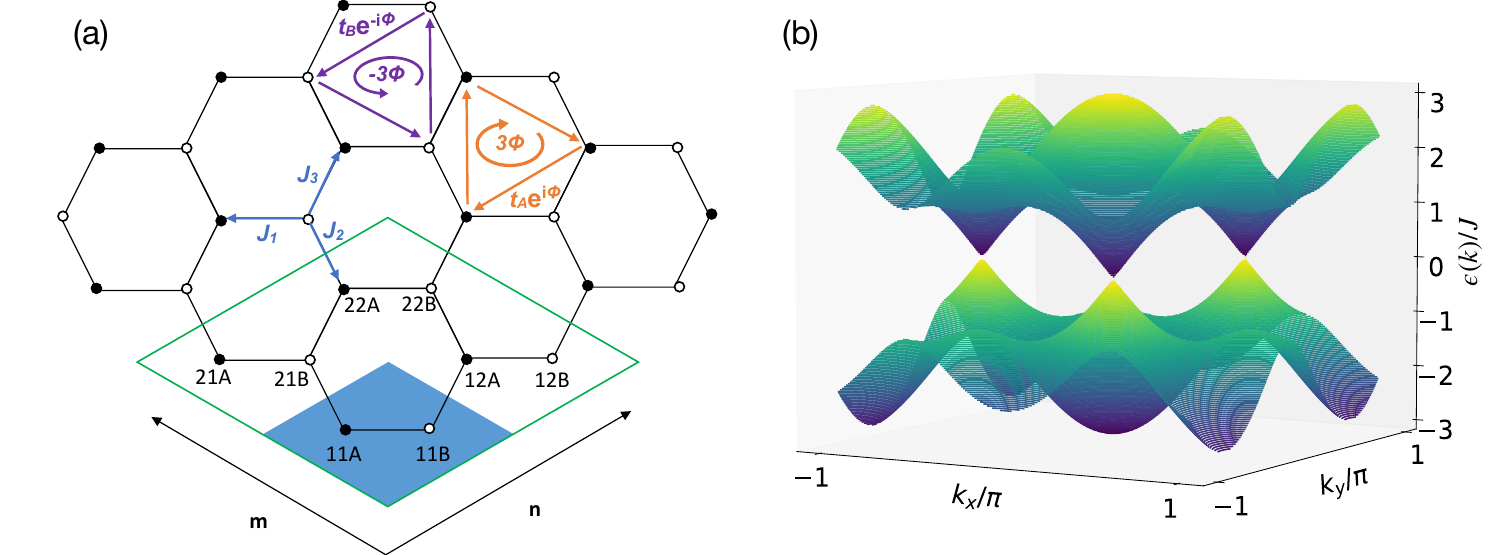}
	\caption{(a) Haldane model with hopping vectors in real space. The region in blue shading shows one unit cell with two sites $A$ and $B$. $J_1$, $J_2$ and $J_3$ are the nearest neighbour hopping amplitudes. The next nearest neighbour (NNN) hopping in a clockwise closed path with magnetic flux $3\phi$ enclosed is shown in orange color and the corresponding NNN hopping acquires a complex phase $e^{i\phi}$, and vice versa, the anticlockwise NNN hopping acquires a complex phase $e^{-i\phi}$. In order to enumerate the unit cells, we consider two directions with labels $m$ and $n$, which take integer values. Each lattice site is described by three parameters: $mn\sigma$, with $mn$ the index of unit cell and $\sigma = A, B$. (b) Single particle spectrum under periodic boundary condition for the Haldane model with $J_\lambda = J$, $t_{A} = t_B = 0.1J$, $\phi=\pi /2$, $\Delta = 0.52J$.}
	\label{fig:haldane_lattice}
\end{figure}

After having demonstrated the effectiveness of the cooling protocols in one-dimensional topological insulators, we now address the case of a two-dimensional system: the Haldane model of a Chern insulator~\cite{Haldane}. This model substantially differs from previous examples, both because of its dimensionality and of its topological characterization. The Haldane model is defined on a honeycomb lattice with two sites ($A$ and $B$) per unit cell, as shown in Fig. \ref{fig:haldane_lattice}(a). In order to enumerate the unit cells, we consider two directions with labels $m$ and $n$, which take integer values. Each lattice site is described by three parameters: $mn\sigma$, with $mn$ the index of the unit cell and $\sigma = A, B$. The Hamiltonian is given by
\begin{flalign}
\label{eq:HaldaneHamiltonian}
	H =& -\sum_{\langle i,j \rangle_{\lambda=1,2,3} } J_\lambda \left( a_i^{\dagger}b_j+h.c. \right) 
    -\sum_{\langle\!\langle i, j \rangle\! \rangle_A} t_A \left(a_i^{\dagger}a_je^{i\nu_{ij}\phi}+h.c.\right) \\ \nonumber
    &-\sum_{\langle\!\langle i, j \rangle\! \rangle_B} t_B \left(b_i^{\dagger}b_je^{i\nu_{ij}\phi}+h.c.\right)
	+\Delta \sum_{i}\left(a_i^{\dagger}a_i-b_i^{\dagger}b_i \right) \ .
\end{flalign}

The summation over $\langle i,j\rangle_\lambda$ runs over (ordered) pairs of nearest neighbours (NN) with $\lambda=1,2,3$, as shown in Fig. \ref{fig:haldane_lattice}(a), while the summation over $\langle\!\langle i,j\rangle\!\rangle_{A/B}$ runs over next-nearest neighbours (NNN) between $AA$ or $BB$ sites. The complex phases $e^{i\nu_{ij} \phi}$ can be thought of as resulting from a magnetic field penetrating the lattice with zero net flux in each hexagon. The value of $\nu_{ij}$ is determined by the hopping directions with $\nu_{ij}=1$ for clockwise hopping and $\nu_{ij}=-1$ for counterclockwise hopping. The Haldane Hamiltonian thus features real-valued NN hopping parameters ($-J_\lambda$) and complex NNN hopping parameters ($-t_\sigma e^{i\nu_{ij}\phi}$). We consider first the original Haldane model with $J_\lambda=J$ and $t_\sigma = t$, if not otherwise mentioned.
The complex NNN coupling matrix elements break time-reversal symmetry and open gaps at both of the Dirac-type band-touching points, so that the resulting individual bands require topologically non-trivial properties characterized by Chern numbers $\pm 1$. In turn, the energy offset $\Delta$ breaks inversion symmetry. This term alone would open a topologically trivial band gap. If both terms are present, we find a competition between the two. By varying $\Delta$, the system enters a topological phase when $|\Delta|<|3\sqrt{3}t\sin\phi|$, which is characterized by the appearance of the chiral edge states\ \cite{PhysRevB.74.235111}.

\subsubsection{Exact vs approximate method in small systems}\label{small-system}
\begin{figure}
\centering
    \includegraphics[width=0.5\linewidth]{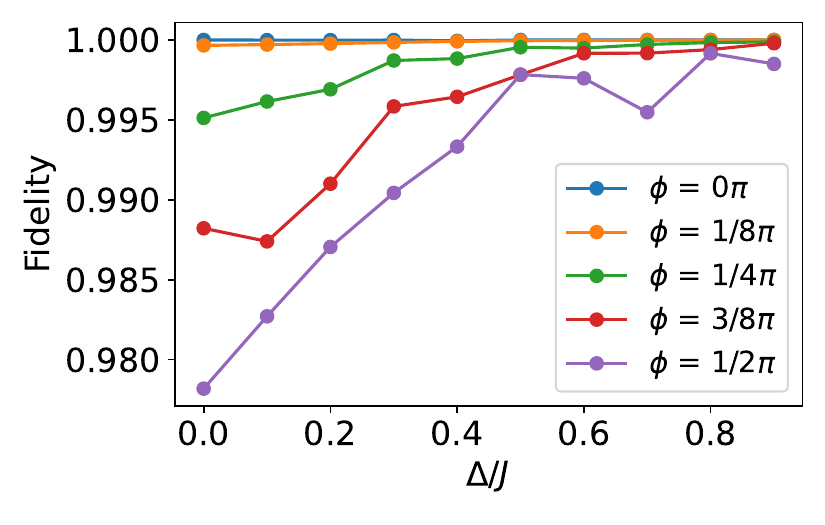}
	\caption{(a) Fidelity between the steady state given by Eq.~\eqref{me_fbM} with jump operator $C$ of Eq.~\eqref{eq:jump_exact} (for the exact approach) and the ground state of the Haldane model~\eqref{eq:HaldaneHamiltonian} for a small system with four unit cells as a function of the on-site potential $\Delta$ for different values of the complex hopping phase $\phi$. The steady state is obtained by solving Eq.~\eqref{me_fbM}. We use OBC with a small interaction~\eqref{eq:Hint} of strength $0.001J$ to break the degeneracy.
    }%
	\label{fig:Haldane_intermediate_state}%
\end{figure}
%
To investigate feedback cooling in the Haldane model, we first focus on a system with $N=4$ unit cells as depicted with the green line in Fig. \ref{fig:haldane_lattice}(a).  
The steady state is obtained by solving Eq.~\eqref{me_fbM} as done for the  Rice-Mele model. Unwanted degeneracies arising from the symmetric single-particle spectrum shown in Fig.~\ref{fig:haldane_lattice}(b) are lifted by using open boundary conditions and adding a small nearest-neighbour interaction, see Eq.~\eqref{eq:Hint}. The fidelity given by the exact method is shown in Fig.~\ref{fig:Haldane_intermediate_state}a for different values of the NNN tunneling phase $\phi$. As for the 1D case, the fidelity is very close to one for all parameter values, confirming the efficiency of the exact method.  
We further observe that the fidelity deviates slightly from one for smaller $\Delta$ and larger $\phi$: we attribute this discrepancy to the fact that, for the value $U=0.001J$ of the interaction strength of Eq.~\eqref{eq:Hint} used, the system exhibits an increasing number of near-degeneracies in this regime, with decreasing level spacing as $\Delta$ decreases and $\phi$ increases. These near-degeneracies hinder the efficiency of the exact protocol through the phenomena discussed in Sec.~\ref{basic-idea}.
\begin{figure}[ht]
	\includegraphics[trim = 5mm 60mm 5mm 69mm, clip, width=1
    \textwidth]{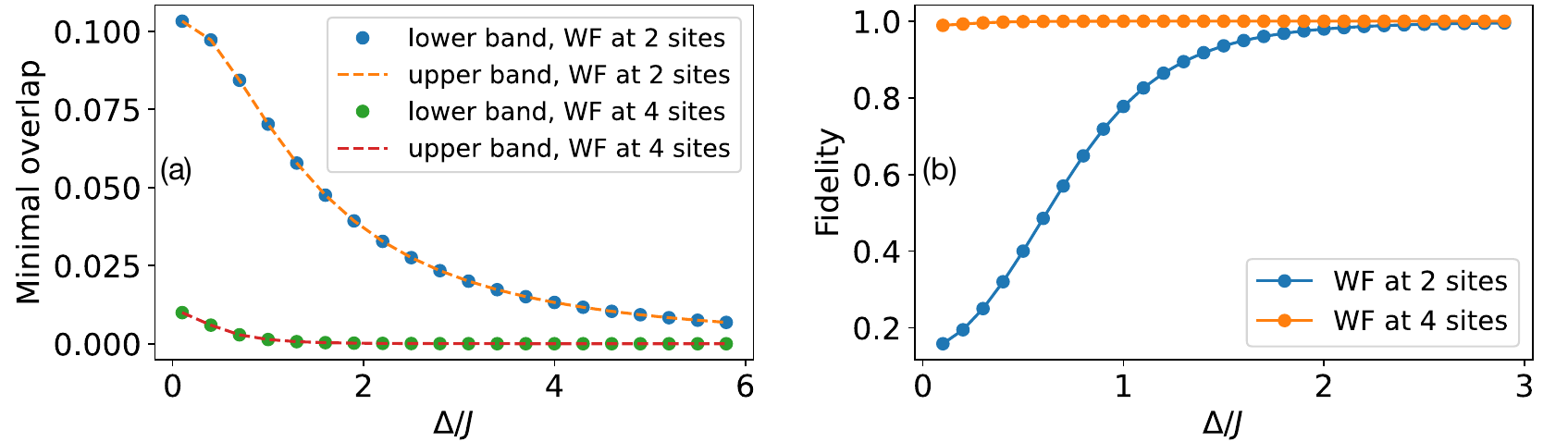}
	\caption{Approximate approach for a small system of the Haldane model with 4 unit cells as depicted with green line in Fig. \ref{fig:haldane_lattice}(a). We consider the cases where the wavefunctions (WF) $\ket{b_\pm^{\mathcal{S}_n}}$ are localized at two sites with indices $11A$, $11B$ in the first unit cell, and on four sites with indices $11A$, $11B$, $21B$, $12A$ indicated in Fig.~\ref{fig:haldane_lattice}(a). We set $\phi=\pi/2$ and $t = 0.1J$, where the energy unit $J$ is the NN hopping amplitude. The parameterizations of $\ket{b_\pm^{\mathcal{S}_n}}$ are given by Eq.~\eqref{ansatz} for $n=2$ and by Eq.~\eqref{ansatz_4sites} for $n=4$. The parameters can be determined by minimizing the overlap of $\ket{b_-^{\mathcal{S}_n}}$ ($\ket{b_+^{\mathcal{S}_n}}$) with upper (lower) band, with the minimal overlaps shown in (a).
	(b) Fidelity between the steady state given by Eq.~\eqref{me_fbM} and the ground state of the Haldane model for a small system with 4 unit cells as a function of the onsite potential $\Delta$ for $t=0.1J$. The steady state is obtained by solving Eq.~\eqref{me_fbM} with the jump operator $C_{\mathcal{S}_n}= (b_-^{\mathcal{S}_n})^\dagger b_+^{\mathcal{S}_n}$. The measurement strength is set to $\gamma=0.0001J$.
	}%
	\label{fig:Haldane_approach2}%
\end{figure}
\begin{figure}[t]
	\includegraphics[trim = 30mm 80mm 35mm 28mm, clip, width=1
    \textwidth]{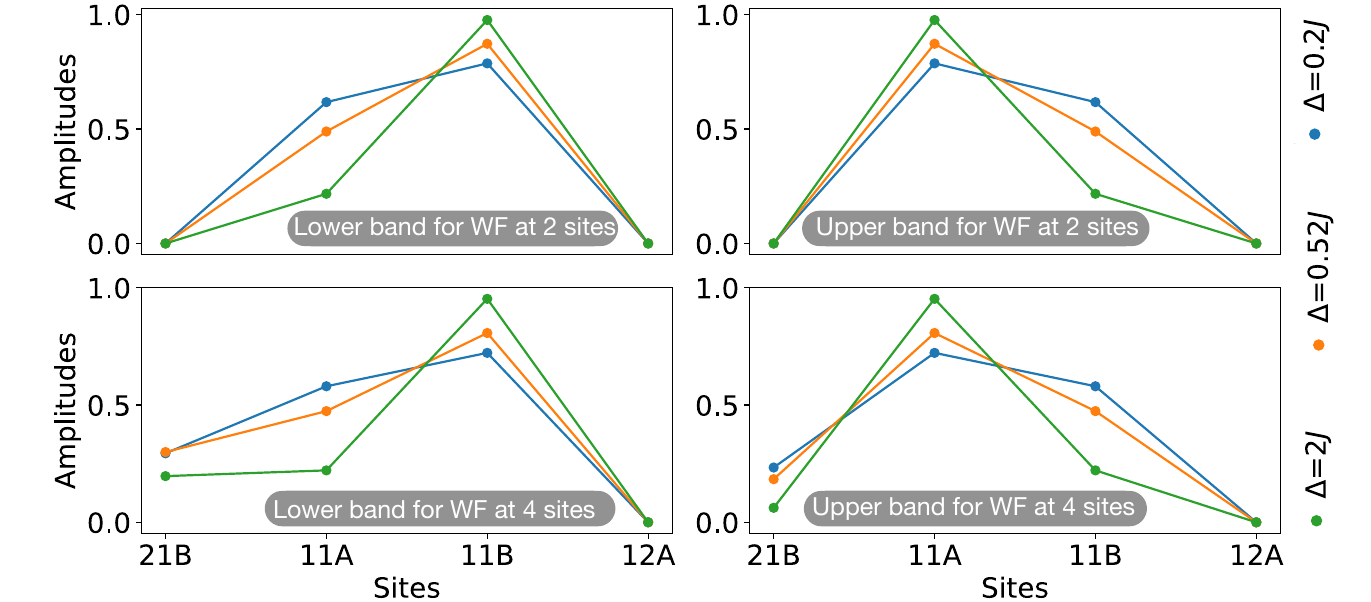}
	\caption{Amplitudes of the optimal wavefunctions $\ket{b_{\ell=1,\pm}}$ at two sites in the first unit cell with the ansatz given by Eq.~\eqref{ansatz} and $\ket{b_{\pm}^{\mathcal{S}_4}}$ at four sites with the ansatz given by Eq.~\eqref{ansatz_4sites}. The different colors stand for different on-site potentials $\Delta$. We consider a small system of the Haldane model with 4 unit cells as depicted with green line in Fig. \ref{fig:haldane_lattice}(a). We set $\phi = \pi/2$ and $t=0.1J$ with $J$ the NN hopping amplitude. To get the optimal parameters we try to minimize the overlap of $\ket{b_-^{\mathcal{S}_n}}$ ($\ket{b_+^{\mathcal{S}_n}}$) with upper (lower) band. We choose the two sites with indices $11A$, $11B$ in the first unit cell, and the four sites with indices $11A$, $11B$, $21B$, $12A$, as shown in Fig.~\ref{fig:haldane_lattice}(a). 
	}%
	\label{fig:Haldane_amplitudes}%
\end{figure}

We then test the approximate method by constructing constrained jump operators localized at two sites with indices $11A$, $11B$ in the first unit cell, and on four sites with indices $11A$, $11B$, $21B$, $12A$ indicated in Fig.~\ref{fig:haldane_lattice}(a). We can parameterize the wavefunctions localized at four sites with three parameters $\xi_1$, $\xi_2$ and $\xi_3$ in the following way,
\begin{align}\label{ansatz_4sites}
  \ket{b_{\pm}^{\mathcal{S}_4}} = &\sin(\xi_1) \sin(\xi_2) \sin(\xi_3) \ket{11A} + \sin(\xi_1) \sin(\xi_2) \cos(\xi_3) \ket{11B} \nonumber \\
  & + \sin(\xi_1) \cos(\xi_2)\ket{21B}+\cos(\xi_1)\ket{12A}.
 \end{align}
Given the parametrization \eqref{ansatz_4sites}, the optimal parameters $\xi_1$, $\xi_2$ and $\xi_3$ can be determined by following a similar strategy as for the two-sites case, where we try to minimize the overlap of $\ket{b_-^{\mathcal{S}_4}}$ ($\ket{b_+^{\mathcal{S}_4}}$) with the upper (lower) band. The minimal overlaps found are shown in Fig.~\ref{fig:Haldane_approach2}(a). For smaller on-site potential $\Delta$ (corresponding to the topological phase) the minimal overlaps are smaller for $\ket{b_\pm^{\mathcal{S}_n}}$ localized at four sites than at two sites. The amplitudes of the optimal wavefunction $\ket{b_\pm^{\mathcal{S}_n}}$ at different lattice sites are shown in Fig.~\ref{fig:Haldane_amplitudes}. 
The resulting fidelity is shown in Fig.~\ref{fig:Haldane_approach2}(b). For the two-site jump operator $C_{\mathcal{S}_2}$, the cooling is not efficient for small on-site potentials $\Delta/J\ll 1$, but the fidelity grows monotonically by increasing $\Delta$, eventually reaching $\mathcal{F}=1$ for $\Delta/J\sim 3$. The situation strikingly improves when considering $C_{\mathcal{S}_4}$, localized on four sites. In this case the fidelity barely deviates from one at vanishing $\Delta$, and remains otherwise very close to one for all values of $\Delta$. 
For small on-site potentials, the 2D system is in a topological phase and it is not possible to construct maximally localized Wannier functions\ \cite{PhysRevB.74.235111}, which leads to poor cooling performance of the two-site algorithm. This differs significantly from the 1D SSH model \ref{cooling_ssh}, where cooling in the topological phase still works by simply shifting the operators $\ket{b_{\ell,\pm}}$, as the maximally localized Wannier functions are shifted in this case.

\subsubsection{Cooling large systems: mean-field approach}\label{mean-field}

Having benchmarked the performance of both the exact and the approximate methods in small systems by exact numerical determination of the steady state, we now investigate cooling in the (non-interacting) Haldane model for larger systems by adopting a mean-field approach. In particular, following Refs.~\cite{PhysRevLett.111.240405, vorberg} we derive kinetic equations of motion for the mean occupations of the single-particle eigenstates (\ref{section:mean_field_derivation}). The resulting non-linear equations of motions read
\begin{equation} \label{eq:mfeq}
\frac{d}{dt}\bar{n}_k(t) = \sum_q\Big( R_{kq}\bar{n}_q[1-\bar{n}_k(t)]  -R_{qk}\bar{n}_k(t)[1-\bar{n}_q(t)]\Big),
\end{equation}
where $\bar{n}_k(t)=\mathrm{tr}[n_k \rho(t)]$ is the mean occupation number of the $k$-th single-particle eigenstate (ordered by increasing energy). These equations are obtained by neglecting non-trivial correlations, $\mathrm{tr}[n_kn_q\rho(t)]\approx \bar{n}_k(t)\bar{n}_q(t)$, which is justified in the limit of large systems. The validity of the mean-field approximation for small systems is further studied in Section \ref{sec:validity_meanfield} in the Appendix, where the exact full-density-matrix approach is compared with mean-field approach. The transition rates $R_{kq}$ are obtained from the feedback master equation \eqref{me_fbM0} after an additional rotating-wave-approximation (RWA), which is valid if the single-particle energy gaps and their differences are much larger than the measurement strength $\gamma$. The rates are derived from the feedback jump operator $C$ according to $R_{kq}=|C_{kq}|^2$.

In the case of the exact method, we see from Eq. \eqref{eq:Abloch} that the rates $R_{kq}$ are such that $R_{kq}=1$ (in the dimensionless units used), if the eigenstates labeled by $k$ and $q$ belong to the lower and upper band, respectively, while $R_{kq}=0$ otherwise. The mean-field equation \eqref{eq:mfeq} for the occupation of a state $k$ in the lower band then reduces to
\begin{equation}
\frac{d}{dt} \bar{n}_k(t) = [1-\bar{n}_k(t)]\sum_{q \in B_+} \bar{n}_q(t) ,
\end{equation}
where $B_+$ denotes the upper band. The steady state, $d\bar{n}_k(t)/dt=0$, is then easily found to be $\bar{n}_k(t)=1$, confirming that an exact preparation of the many-body ground state is indeed attained, featuring all states in the lower band occupied. 

\begin{figure}
\includegraphics[trim = 0mm 82mm 20mm 46mm, clip, width=1
    \textwidth]{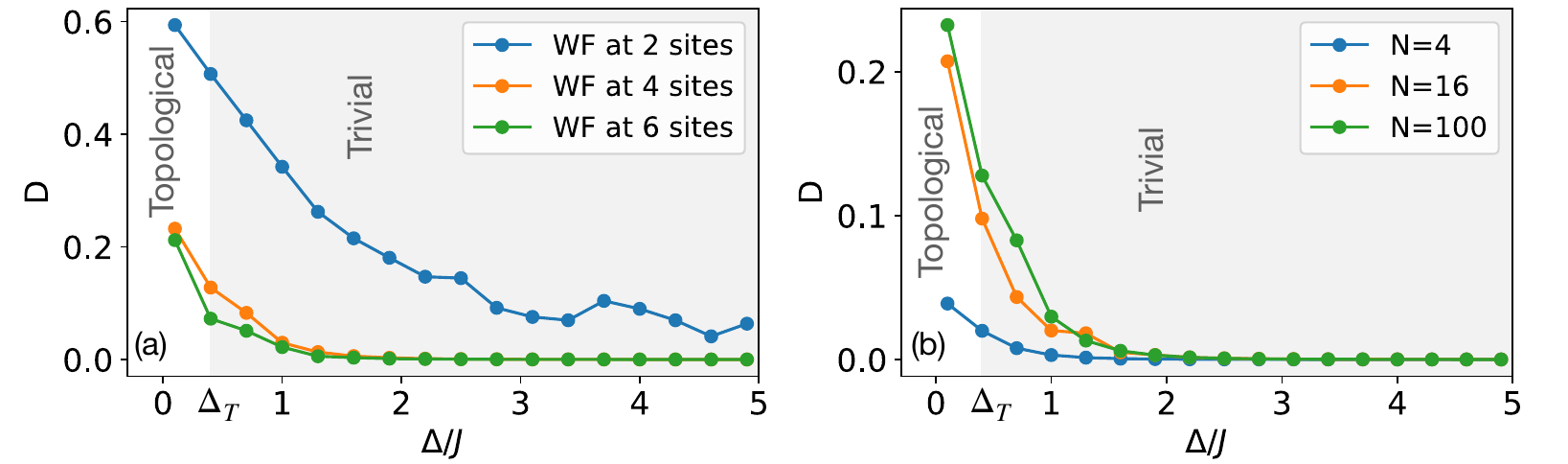}
    \caption{ Efficiency of the approximate construction for the Haldane model. We plot the deviation given by~\eqref{eq:Deviation} between the steady state of the mean-field equations~\eqref{eq:mfeq} and the ground state as a function of on-site potential $\Delta$ for $\phi=\pi/2$. We consider slightly different NNN hoppings with $t_A=0.1J$, $t_B=0.05J$, where $t_A$ denotes the NNN hopping amplitude without phase factor for $AA$ sites and $t_B$ denotes the NNN hopping amplitude without phase factor for $BB$ sites. $\Delta_T$ denotes the topological phase transition. (a) We consider a large system with 100 unit cells arranged in $10\times 10$ , as shown in Fig. \ref{fig:haldane_lattice}(a). The wavefunctions are constrained at two, four or six sites . The choice of $\mathcal{S}_2$ and $\mathcal{S}_4$ is same with Fig. \ref{fig:Haldane_approach2}. $\mathcal{S}_6$ is given by $11A$, $11B$, $21B$, $12A$, $22A$ and $22B$ as indicated in Fig.~\ref{fig:haldane_lattice}(a). (b) Deviation for different large systems with $N$ unit cells arranged in a manner such that $m$ and $n$ run from 1 to $\sqrt{N}$, as shown in Fig. \ref{fig:haldane_lattice}(a). The wavefunctions are constrained at four sites with the choice of $\mathcal{S}_4$ same with (a).}
    \label{fig:haldane_100}%
\end{figure}
\begin{figure}[ht]
	\centering
	\includegraphics[trim = 45mm 71mm 35mm 40mm, clip, width=0.8
    \textwidth]{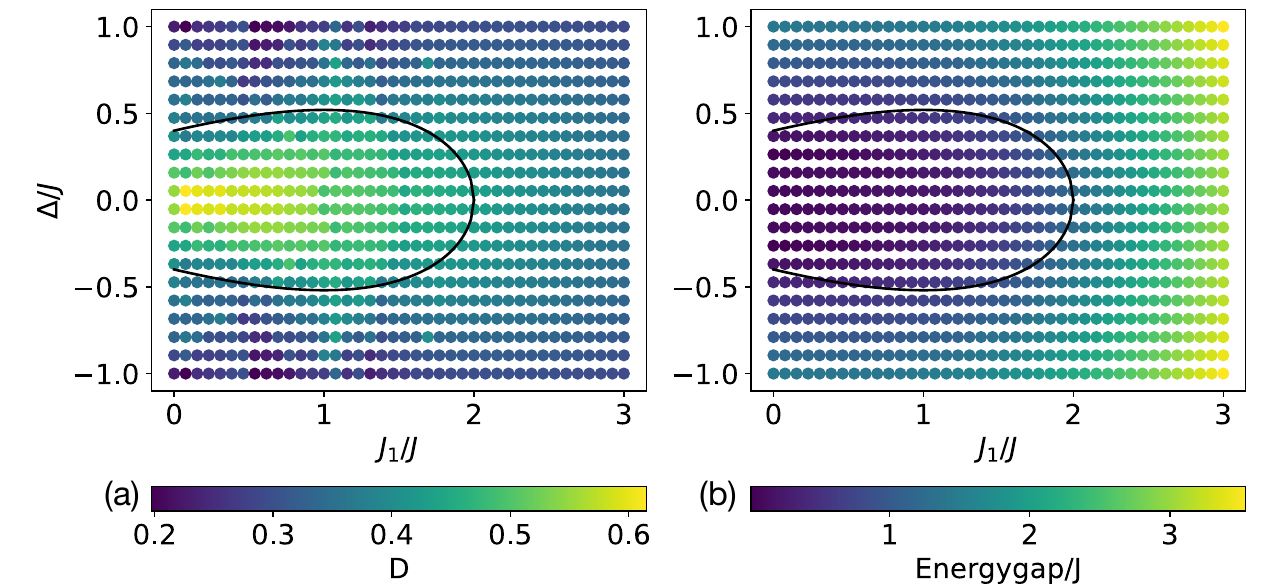}
	\caption{Deviation (a) between final state of the mean-field equations (\ref{eq:mfeq}) and the ground state of the Haldane model for a large system with 16 unit cells arranged in the same way as Fig. \ref{fig:haldane_100}. (b) the corresponding energy gaps under open boundary conditions. The approximate approach with wavefunctions localized at 2 sites is applied here. The parameters are set as $J_2=J_3=J$ and $t=0.1J$. The black curves describe the topological transition, which is given by Eq. (\ref{eq:Hz_dimerization}). The horizontal axis is $J_1$ and vertical axis is the on-site potential $\Delta$. The deviation $D$ given by (\ref{eq:Deviation}) and the energy gap, which are functions of $J_1$ and $\Delta$, are represented with a color map.}%
	\label{fig:haldane_gap}
\end{figure}
We then investigate the performance of the approximate method by constraining the jump operator to different sets of sites $\mathcal{S}_n$ and solving the mean field equations \eqref{eq:mfeq} numerically in a system of $N=100$ unit cells arranged in a manner such that $m$ and $n$ run from 1 to 10, as shown in Fig. \ref{fig:haldane_lattice}(a). To ensure that the spectrum does not feature two identical level spacings (see \ref{section:mean_field_derivation}), such that the RWA can be assumed to be valid for a suitably small $\gamma$, we consider slightly different strength of the NNN hopping between two $A$ sites, denoted with $t_A$, as compared to two $B$ sites, denoted with $t_B$ (see Fig.~\ref{fig:haldane_lattice}(a)). The construction of the constrained jump operators is performed via numerical optimization as explained in Section~\ref{sec:appr_meth}. We quantify the deviation $D(\bar{\bm{n}})$ from the ground state by comparing the mean occupation numbers in the steady state, $\bar{\bm{n}}=\{\bar{n}_k\}_{k=1,\dots,2N}$ with those in the ground state, $\bar{\bm{n}}^{(g)}=\{\bar{n}_k^{(g)}\}_{k=1,\dots,2N}$, computing
\begin{equation} \label{eq:Deviation}
D(\bar{\bm{n}}) = \sqrt{\frac{1}{2N}\sum_{k=1}^{2N}(\bar{n}_k - \bar{n}_k^{(g)})^2}.
\end{equation}
The sets $\mathcal{S}_n$ considered are localized on $n=2$, 4 and 6 sites. The choice of $\mathcal{S}_2$ and $\mathcal{S}_4$ is the same as in Fig.~\ref{fig:Haldane_approach2}. $\mathcal{S}_6$ is given by $11A$, $11B$, $21B$, $12A$, $22A$ and $22B$ as indicated in Fig.~\ref{fig:haldane_lattice}(a). 
The Deviation $D$ obtained as a function of $\Delta$ is depicted in Fig.~\ref{fig:haldane_100}. We observe that the deviation is fairly large for $\mathcal{S}_2$~(blue curve), but it gets much closer to zero as the number of sites is increased, for $\mathcal{S}_4$~(orange curve) and $\mathcal{S}_6$~(green curve). In the latter cases, the deviation tends to zero for $\Delta/J>1$, but deviates from zero for $\Delta/J<1$. 

The above effect can be understood in the light of the topological phase transition, thus interestingly linking the efficiency of the engineered cooling mechanism with the system's topological properties. Indeed, around $0<\Delta_T<1$, the system crosses the phase transition and enters the topological phase, where exponentially localized Wannier functions do not exist, such that the constrained ansatz states $\ket{b_\pm^{\mathcal{S}_n}}$ cannot reproduce the eigenstates to a satisfactory degree. To bring further evidence that this effect is indeed related to the topological phase transition, we show in Fig.~\ref{fig:haldane_100}(b) the deviation for different system sizes, again as a function of $\Delta$. The steepness of the curve at low $\Delta$ increases for increasing size, as expected, since it is more difficult to find Wannier functions localized at four sites in the topological phase.
As an additional signature, we study how the cooling improves by entering a trivial dimerized phase, in which the tunneling strength $J_1$ along the direction $e_1$ is much larger than in other directions~[see Fig.~\ref{fig:haldane_lattice}(a)]. For the case of a dimerization with changing $J_1$ and constant $J_2=J_3=J$ the topological phase transition is given by
\begin{equation} \label{eq:Hz_dimerization}
    \Delta \pm t \left(\frac{J_1}{J}+2\right)\sqrt{4-\frac{J_1^2}{J^2}} = 0.
\end{equation}
The deviation for this case is reported in Fig.~\ref{fig:haldane_gap}(a), and indeed becomes smaller for increasing $J_1$. This observation together with the results in Fig.~\ref{fig:haldane_100} 
indicate that the deviation is related with the energy gap, so we plot the energy gaps of the Haldane model in Fig.~\ref{fig:haldane_gap} for the approximate approach with jump operators localized at two sites under open boundary condition. We observe that $D$ is smaller (larger) for a larger (smaller) energy gap, which indicates that the cooling performance is much better for a larger energy gap.

\section{Conclusions}	\label{conclusion}
In summary, we have proposed a scheme to prepare topological insulator states for noninteracting fermionic atoms in an optical lattice through Markovian feedback control. Specifically, we considered topologically non-trivial two-band models at half-filling, and exploited continuous weak measurement and Markovian feedback to engineer a dissipative process which cools the system towards the ground state. This is achieved, in turn, by constructing a dissipative process that pumps particles from the upper band to the lower band, until the latter is filled. We further propose approximate variant schemes that can perform the same task with lower efficiency, when additional experimental constraints on the measurement and feedback apparatus are introduced. We have benchmarked these two approaches in several 1D and 2D lattice models, namely for the Su-Schrieffer-Heeger model, the Rice-Mele model and the Haldane model. For moderate system sizes, we probed the steady state of the system by solving the feedback-modified master equation numerically. For large systems, we resorted to kinetic theory and compared the mean occupations of the single-particle eigenstates. The proposed exact cooling scheme is successful in all parameter regimes and for all models studied. The approximate methods, which involve a restriction of the measurement and feedback operations to small subsystems, give good performance for small systems or when the system eigenstates tend to be localized on few sites. While this makes this approach less effective in the topological phase of the 2D Haldane model for large systems, it still gives a satisfactory preparation of TI states in the 1D models studied.

\section*{Acknowledgements}

\paragraph{Funding information}
This research was funded by the Deutsche Forschungsgemeinschaft (DFG, German Research Foundation) via the Research Unit FOR 5688 “Driven-dissipative many-body systems of ultracold atoms”, project number 521530974, and by the Hainan Province Science and Technology Talent Innovation Project (Grant No.~KJRC2023L05). F.~P. acknowledges funding from the Deutsche Forschungsgemeinschaft (DFG, German Research Foundation) through the Emmy Noether Programme -- project number 555842149.

\begin{appendix}

\section{Impact of measurement strength on fidelity}
\label{section:change_gamma}
\begin{figure}[ht]
\centering
		\includegraphics[trim = 0mm 96mm 0mm 30mm, clip, width=1\textwidth]{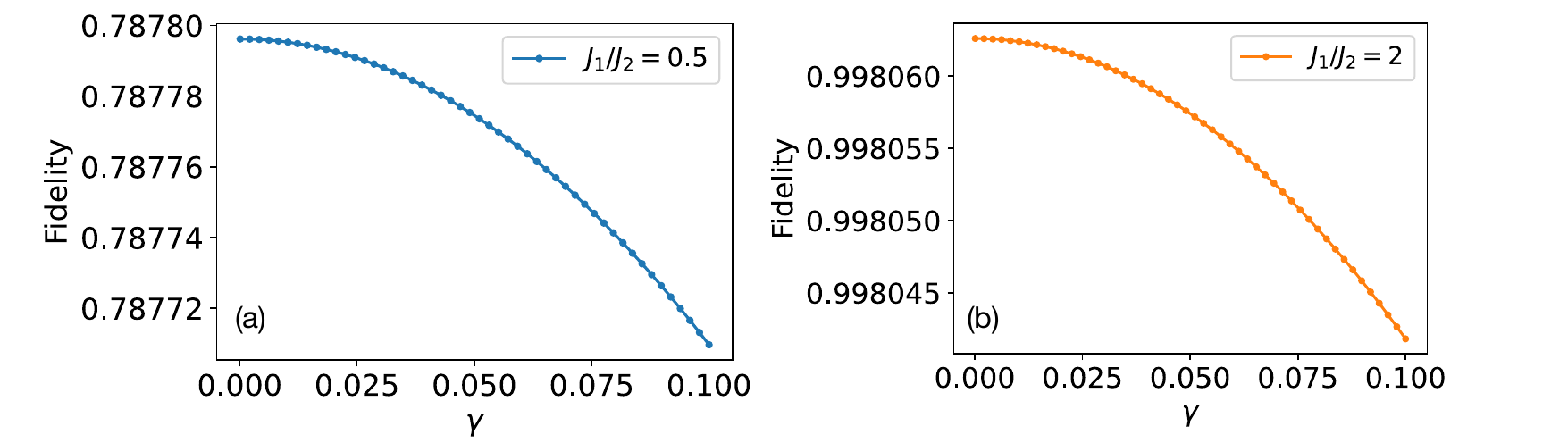}
        \caption{Fidelities for a small SSH model with 3 unit cells for different values of measurement strength $\gamma$ by considering two different topological regimes (a) topological phase and (b) trivial phase, for the approximate approach with Wannier function localized in one unit cell.}
        \label{fig:fidelity_diff_gamma}
\end{figure}

For the simulations in the paper, we did not include the feedback term $H_{\text{fb}}$. To verify that this approximation does not alter the results significantly and to address the impact of the measurement strength $\gamma$ on the steady-state fidelity, we have added $H_{\text{fb}}$ to the system Hamiltonian and computed fidelities for a small SSH model with three unit cells, considering different values of $\gamma$ and two different topological regimes: $J_1/J_2=2$ and $J_1/J_2=0.5$, as shown in Fig.~\ref{fig:fidelity_diff_gamma}. The setup is the same as in Fig.~\ref{fig:sshapproachII_fidelitydifferentunitcellnumbers} for the approximate approach. By slowly changing $\gamma$ from 0.0001 to 0.1, we observe that a smaller $\gamma$ gives slightly higher fidelity, but in both cases the influence of $\gamma$ in fidelity is on the order of $10^{-5}$, thus justifying the omission of the $H_{\text{fb}}$ term in the simulations.

\section{Different methods to lift degeneracies of the many-body spectrum}
\label{section:lift_degeneracy}
\begin{figure}[ht]
\includegraphics[trim = 35mm 61mm 48mm 35mm, clip, width=1
    \textwidth]{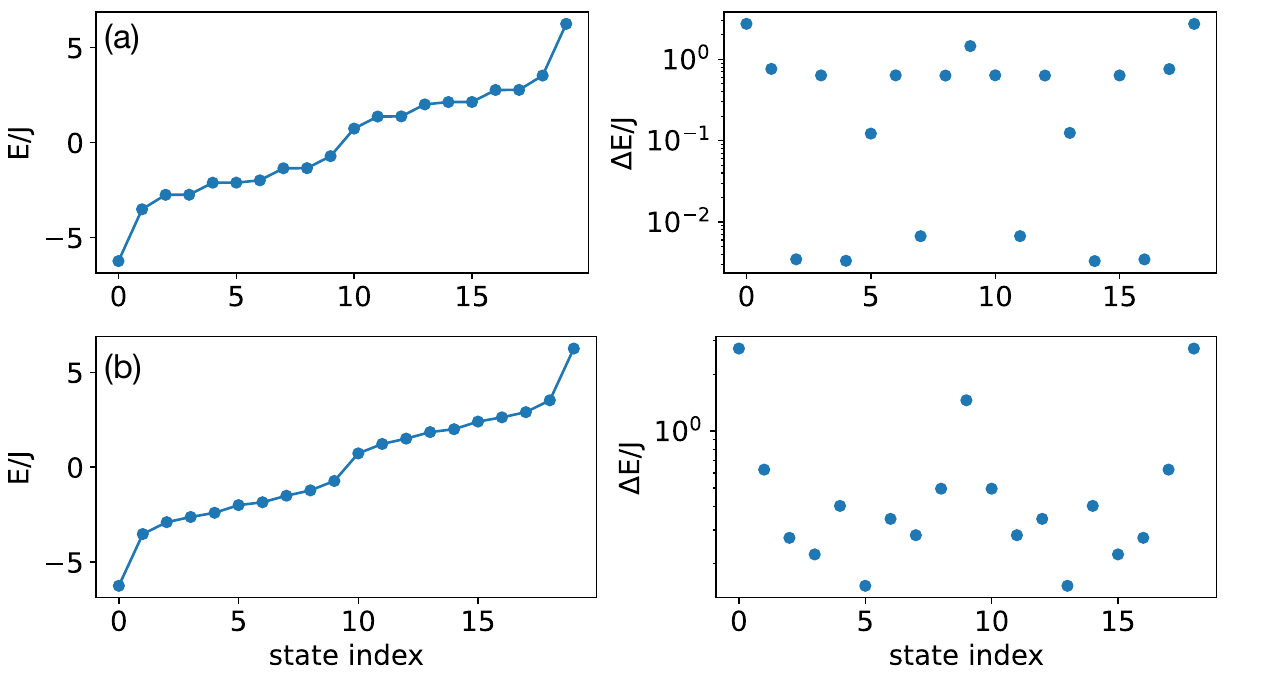}
    \caption{Half-filling spectrum for a system of 3 unit cells in SSH model with $J_1=2J$ and $J_2=J$. The eigenenergies are sorted in an ascending way and enumerated along the x-axis. Figures on the left side are plots of the energy spectrum in linear scale. Figures on the right side are in semilog
scale and describe the energy difference between neighboring points of the left plots. (a) Plots for the
case of OBC with small interaction of strength 0.01J. (b) Plots for the
case of OBC with NNN hopping strength 0.1J.}
    \label{fig:breakDegeneracy}%
\end{figure}
The degeneracies in the many-body spectrum arise from the symmetric single-particle spectrum with respect to $\pm k$ and between the two bands, as shown in Fig. \ref{fig:rice_mele_ssh_potential}(b). We study different ways to lift the degeneracies.\\
(1) We introduce a small interaction between nearest neighbors with interaction Hamiltonian in Eq.~\eqref{eq:Hint}, with $U$ denoting the interaction strength, and use open boundary conditions (OBC). For OBC $l$ runs from 1 to $N-1$ in Eq.~(\ref{eq:rice_mele_Hamiltonian}). We plot the half-filling spectrum and the energy difference between neighbouring states in Fig.~\ref{fig:breakDegeneracy}(a), where all the energy differences $\Delta E$ are non-zero, which means that the degeneracy is lifted by adding small nearest-neighbour interactions.\\
(2) We can introduce more tunneling coefficients, such as next nearest neighbor tunneling coefficients (NNN hopping). Similar as in (1), the half-filling spectrum and the energy differences are plotted in Fig.~\ref{fig:breakDegeneracy}(b).\\
\section{Maximally localized Wannier function in SSH model}
\label{section:MLWF}
\begin{figure}
\includegraphics[trim = 0mm 5mm 0mm 60mm, clip, width=0.98\textwidth]{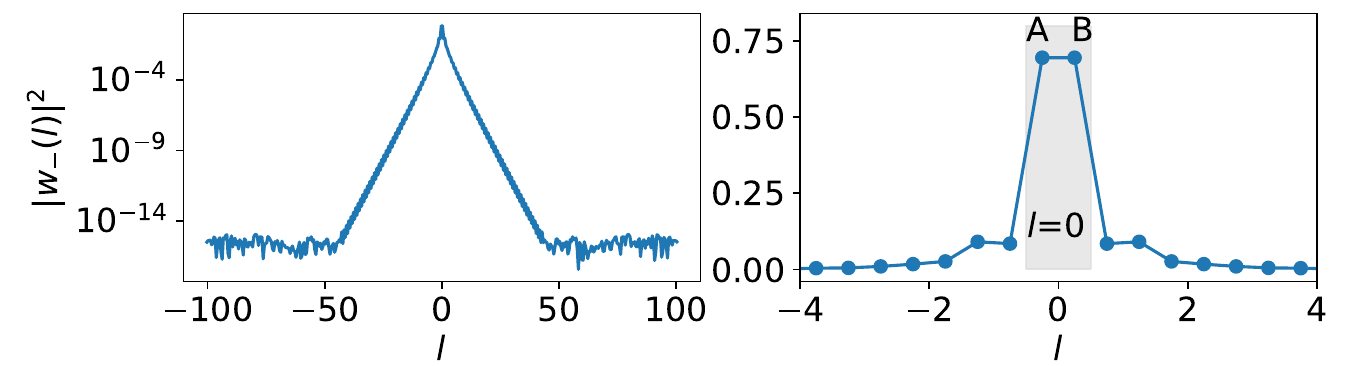}
    \caption{MLWF in tight binding model for SSH model with 201 unit cells with $J_1=2J_2$, for the lower ($-$) band. The index $l$ lists all unit cells from -100 to 100. The gray region shows the 0th unit cell with $A$ and $B$ sites. In this case the Wannier function is exponentially localized. Plots on the left side are in semilog scale. Plots on the right side are in linear scale and zoom on the peak of the Wannier functions.}
\label{fig:MLWF_ssh}%
\end{figure}
We recall that Wannier functions are used to construct the collapse operators in Eq.~\eqref{eq:Abloch}. The approximate approach, where the collapse operator is chosen to be localized at a few sites, is based on the fact that by choosing appropriate gauge factors it is possible to find well localized Wannier states in Eq.~\eqref{eq:Abloch}. Here, as an example, we try to construct maximally localized Wannier function $\ket{W_{-}(l)}$ in the tight binding SSH model with $J_1=2J_2$, for the lower ($-$) band, where $l$ lists all unit cells. In order to find the appropriate gauge factor $e^{i\varphi_{k_{-}}}$, we perform either a single-band transformation following Ref.~\cite{Modugno} or use the Kohn gauge \cite{Kohn}. In the Kohn gauge, the gauge factor can be chosed as 
\begin{equation} \label{eq:kohnGauge}
 \varphi_{k_{-}}  = -{\rm arg} (\ket{k_{-}}_1+\ket{k_{-}}_2),
\end{equation}
where $\ket{k_{-}}_1$ and $\ket{k_{-}}_2$ are the two components of the lower band eigenvector $\ket{k_{-}}$ in Eq.~\eqref{eq:eigenVector_ssh}. The computed Wannier function is shown in Fig.~\ref{fig:MLWF_ssh}, where we notice an exponential localization of the Wannier function in space. We can calculate the weight of the two sites $0A$ and $0B$ in the 0th unit cell, which is $|w_{-}(0A)|^2+|w_{-}(0B)|^2=0.97$, where the Wannier function is normalized. This result implies that the approximate approach for SSH model should work very well for $J_1=2J_2$, which is consistent with the results in Fig.~\ref{fig:sshapproachII_fidelitydifferentunitcellnumbers}.  
\section{Derivation of the mean-field equations}
\label{section:mean_field_derivation}
Since it is very demanding to calculate the steady state $\rho_{ss}$ numerically for large systems, we can adopt a mean-field description in which we rather study the time evolution of the mean occupation of single-particle eigenstates, following Refs.~\cite{PhysRevLett.111.240405, vorberg}. To work out this mean-field description, we first need to recast the master Eq.~\eqref{me_fbM0} (with $H_{\rm fb}=0$) in a form which describes quantum jumps between single-particle eigenstates. This can be achieved, while maintaining Lindblad form, as follows. We first represent the collapse operator $C$ in the system's eigenbasis, $C = \sum_{k,q} C_{kq} a_k^\dag a_q = \sum_{k,q} C_{kq} L_{kq}$, where  $L_{kq}=a_k^\dag a_q$ describes a quantum jump from single-particle eigenstate $\ket{q}$ to $\ket{k}$. In interaction picture, the master equation then reads as
\begin{align} \label{eq:meapp}
	\frac{d\rho}{dt} = \sum_{k,q,k',q'}C_{kq}C_{k'q'}^* e^{i(\omega_{kq}-\omega_{k'q'})t}\Big[L_{kq} \rho L_{k'q'}^\dag - \frac{1}{2} L_{k'q'}^\dag L_{kq}\rho - \frac{1}{2} \rho L_{k'q'}^\dag L_{kq}\Big],
\end{align}
where $\omega_{kq}=\epsilon_k-\epsilon_q$ is the energy difference between the single-particle levels $\epsilon_k$ and $\epsilon_q$. 
Note that the master equation in Eq.~(\ref{eq:meapp}) is already in Lindblad form, as it is derived from Eq.~(\ref{me_fbM0}), which preserves complete positivity. To further derive the kinetic equations, 
we next adopt the rotating-wave (secular) approximation, which amounts to neglect oscillating terms in Eq.~\eqref{eq:meapp}. This is justified only in the regime where the difference in level spacings $\omega_{kq}-\omega_{k'q'}$ is much larger than the measurement strength $\gamma$, $|\omega_{kq} - \omega_{k^{'}q^{'}}| \gg \gamma$. To ensure that this condition can be met, in our numerical studies we use the methods described in \ref{section:lift_degeneracy} to prevent the emergence of identical level spacings in the single-particle spectrum. In rotating-wave approximation, Eq.~\eqref{eq:meapp} then becomes 
\begin{align} \label{eq:merwa}
	\frac{d\rho}{dt} = \sum_{k,q}R_{kq}\Big[L_{kq} \rho L_{kq}^\dag - \frac{1}{2} L_{kq}^\dag L_{kq}\rho - \frac{1}{2} \rho L_{kq}^\dag L_{kq}\Big],
\end{align}
where we introduced the effective quantum jump rates $R_{kq}=|C_{kq}|^2$. 
From Eq.~\eqref{eq:merwa}, one can derive an Eq.\ ruling the evolution of the single-particle occupations $\bar{n}_k=\mathrm{tr}(\rho n_k)$. The latter will also depend on two-particle correlations, $\mathrm{tr}(\rho n_k n_q)$, initiating a hierarchy of Eqs.~for the $n$-particle correlation functions~\cite{vorberg}. The hierarchy can be truncated in a mean-field-like approximation by assuming the factorization of two-particle correlations, $\mathrm{tr}(\rho n_k n_q)\approx \bar{n}_k\bar{n}_q$~\cite{vorberg}. 
This procedure yields non-linear mean-field equations for the single-particle occupations, reading as
\begin{equation}
	\label{eq:meanfieldequations}
	\frac{d}{dt}\bar{n}_k(t)\approx\sum_{q}\left\{  
	R_{kq}\bar{n}_q(t)[1-\bar{n}_k(t)]-R_{qk}\bar{n}_k(t)[1-\bar{n}_q(t)]
	\right\}.
\end{equation}
In the main text, we studied the steady-state occupations given by these mean-field equations, satisfying $\frac{d}{dt}\bar{n}_k(t)=0$, which have been found numerically both via long-time propagation and through a nonlinear-equation solver.

\section{Validity of the mean-field approximation}
\label{sec:validity_meanfield}

\begin{figure}[t]
\centering
		\includegraphics[width=1\textwidth]{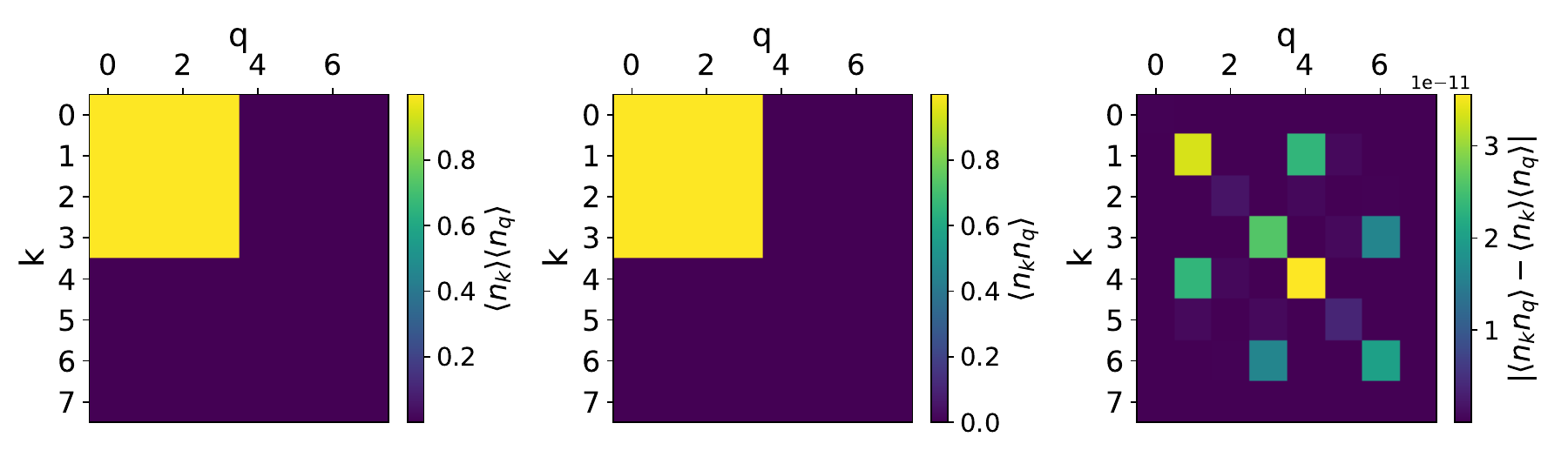}
        \caption{Validity of mean-field approximation for a small SSH system of 4 unitcells with $J_1/J_2=2$: comparison between $\langle n_kn_q \rangle$ and $\langle n_k\rangle \langle n_q \rangle$.}
        \label{fig:compare_nknq1}
\end{figure}

The validity of the mean-field approximation—namely the factorization $\langle n_k n_q\rangle \approx \langle n_k\rangle \langle n_q\rangle$ used in deriving the kinetic equations—is a key assumption for applying our approach to large systems. To assess the reliability of this approximation, we compare the exact results with the mean field results for a small SSH system of four unit cells with $J_1/J_2=2$. We first calculate the steady state $\rho_{ss}$ of the system by solving the master equation. Via $\rho_{ss}$ we compute $\langle n_kn_q \rangle={\rm{tr}}[\rho_{ss} n_k n_q]$ and $\langle n_k\rangle \langle n_q \rangle$, as shown in Fig.~\ref{fig:compare_nknq1}. We observe that $|\langle n_kn_q \rangle-\langle n_k \rangle \langle n_q \rangle|$ is very small, thus justifying the validity of mean-field approximation.

\begin{figure}[t]
\centering
		\includegraphics[trim = 0mm 86mm 0mm 40mm, clip, width=1\textwidth]{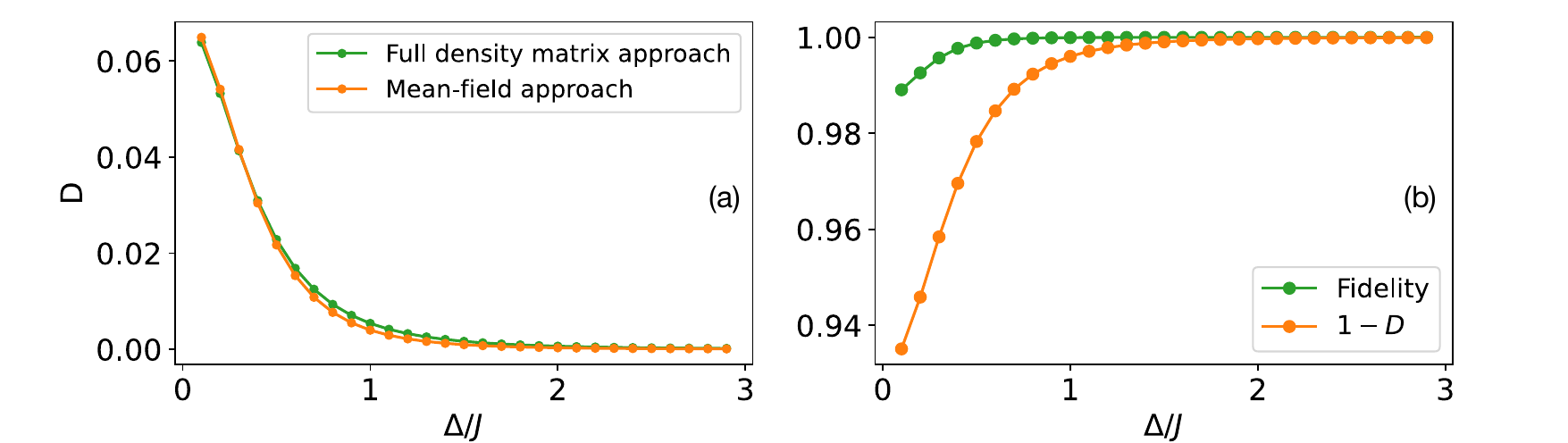}
        \caption{(a) Comparison between the values of the deviation $D(\bar{\bm{n}})$, defined in Eq.~(\ref{eq:Deviation}) of the manuscript, obtained using two different methods: the full steady-state density matrix and the mean-field approach.
        (b) Direct comparison between Fidelity and $1-D$. These studies are done for the small Haldane system with four unit cells, same as in Fig.~\ref{fig:Haldane_approach2}(b) of the main text.}
        \label{fig:Deviation_densitymatrix_meanfield1}
\end{figure}

To further substantiate our approach, we have compared the deviation measure $D$ calculated via both exact diagonalization (ED) and mean-field theory~[Fig.~\ref{fig:Deviation_densitymatrix_meanfield1}(a)]. The observed close agreement between these methods strongly supports the reliability of our mean-field approximation.

Furthermore, we have investigated the relationship between the deviation measure $D$ and the fidelity $\mathcal{F}$. Remarkably, as demonstrated in Fig.~\ref{fig:Deviation_densitymatrix_meanfield1}(b), $1-D$ exhibits behavior qualitatively similar to the fidelity $\mathcal{F}$, suggesting that $1-D$ serves as an excellent proxy for $\mathcal{F}$ in our system.

\end{appendix}



\printbibliography

@article{Johansson_2013,
   title={QuTiP 2: A Python framework for the dynamics of open quantum systems},
   volume={184},
   ISSN={0010-4655},
   url={http://dx.doi.org/10.1016/j.cpc.2012.11.019},
   DOI={10.1016/j.cpc.2012.11.019},
   journal={Comp. Phys. Commun.},
   publisher={Elsevier BV},
   author={Johansson, J.R. and Nation, P.D. and Nori, Franco},
   year={2013},
 pages={1234–1240} }

@article{Doherty1999PRA,
  title = {Feedback control of quantum systems using continuous state estimation},
  author = {Doherty, A. C. and Jacobs, K.},
  journal = {Phys. Rev. A},
  volume = {60},
  issue = {4},
  pages = {2700--2711},
  numpages = {0},
  year = {1999},
  month = {10},
  publisher = {American Physical Society},
  doi = {10.1103/PhysRevA.60.2700},
  url = {https://link.aps.org/doi/10.1103/PhysRevA.60.2700}
}

@article{Wang2001PRA,
  title = {Feedback-stabilization of an arbitrary pure state of a two-level atom},
  author = {Wang, Jin and Wiseman, H. M.},
  journal = {Phys. Rev. A},
  volume = {64},
  issue = {6},
  pages = {063810},
  numpages = {9},
  year = {2001},
  month = {11},
  publisher = {American Physical Society},
  doi = {10.1103/PhysRevA.64.063810},
  url = {https://link.aps.org/doi/10.1103/PhysRevA.64.063810}
}

@article{PhysRevA.71.042309,
  title = {Dynamical creation of entanglement by homodyne-mediated feedback},
  author = {Wang, Jin and Wiseman, H. M. and Milburn, G. J.},
  journal = {Phys. Rev. A},
  volume = {71},
  issue = {4},
  pages = {042309},
  numpages = {9},
  year = {2005},
  month = {04},
  publisher = {American Physical Society},
  doi = {10.1103/PhysRevA.71.042309},
  url = {https://link.aps.org/doi/10.1103/PhysRevA.71.042309}
}

@article{Thomsen2002PRA,
  title = {Spin squeezing via quantum feedback},
  author = {Thomsen, L. K. and Mancini, S. and Wiseman, H. M.},
  journal = {Phys. Rev. A},
  volume = {65},
  issue = {6},
  pages = {061801},
  numpages = {4},
  year = {2002},
  month = {06},
  publisher = {American Physical Society},
  doi = {10.1103/PhysRevA.65.061801},
  url = {https://link.aps.org/doi/10.1103/PhysRevA.65.061801}
}

@article{Campagne2016PRL,
  title = {Using Spontaneous Emission of a Qubit as a Resource for Feedback Control},
  author = {Campagne-Ibarcq, P. and Jezouin, S. and Cottet, N. and Six, P. and Bretheau, L. and Mallet, F. and Sarlette, A. and Rouchon, P. and Huard, B.},
  journal = {Phys. Rev. Lett.},
  volume = {117},
  issue = {6},
  pages = {060502},
  numpages = {5},
  year = {2016},
  month = {08},
  publisher = {American Physical Society},
  doi = {10.1103/PhysRevLett.117.060502},
  url = {https://link.aps.org/doi/10.1103/PhysRevLett.117.060502}
}

@article{Stockton2004PRA,
  title = {Deterministic Dicke-state preparation with continuous measurement and control},
  author = {Stockton, John K. and van Handel, Ramon and Mabuchi, Hideo},
  journal = {Phys. Rev. A},
  volume = {70},
  issue = {2},
  pages = {022106},
  numpages = {11},
  year = {2004},
  month = {08},
  publisher = {American Physical Society},
  doi = {10.1103/PhysRevA.70.022106},
  url = {https://link.aps.org/doi/10.1103/PhysRevA.70.022106}
}

@article{
Geremia2004Science,
author = {JM Geremia  and John K. Stockton  and Hideo Mabuchi },
title = {Real-Time Quantum Feedback Control of Atomic Spin-Squeezing},
journal = {Science},
volume = {304},
number = {5668},
pages = {270-273},
year = {2004},
doi = {10.1126/science.1095374},
URL = {https://www.science.org/doi/abs/10.1126/science.1095374},
eprint = {https://www.science.org/doi/pdf/10.1126/science.1095374}}

@article{Geremia2006PRL,
  title = {Deterministic and Nondestructively Verifiable Preparation of Photon Number States},
  author = {Geremia, JM},
  journal = {Phys. Rev. Lett.},
  volume = {97},
  issue = {7},
  pages = {073601},
  numpages = {4},
  year = {2006},
  month = {08},
  publisher = {American Physical Society},
  doi = {10.1103/PhysRevLett.97.073601},
  url = {https://link.aps.org/doi/10.1103/PhysRevLett.97.073601}
}

@article{Yanagisawa2006PRL,
  title = {Quantum Feedback Control for Deterministic Entangled Photon Generation},
  author = {Yanagisawa, Masahiro},
  journal = {Phys. Rev. Lett.},
  volume = {97},
  issue = {19},
  pages = {190201},
  numpages = {4},
  year = {2006},
  month = {11},
  publisher = {American Physical Society},
  doi = {10.1103/PhysRevLett.97.190201},
  url = {https://link.aps.org/doi/10.1103/PhysRevLett.97.190201}
}

@article{Negretti2007PRL,
  title = {Quantum Superposition State Production by Continuous Observations and Feedback},
  author = {Negretti, Antonio and Poulsen, Uffe V. and M\o{}lmer, Klaus},
  journal = {Phys. Rev. Lett.},
  volume = {99},
  issue = {22},
  pages = {223601},
  numpages = {4},
  year = {2007},
  month = {11},
  publisher = {American Physical Society},
  doi = {10.1103/PhysRevLett.99.223601},
  url = {https://link.aps.org/doi/10.1103/PhysRevLett.99.223601}
}

@Article{Sayrin2011nature,
author={Sayrin, Cl{\'e}ment
and Dotsenko, Igor
and Zhou, Xingxing
and Peaudecerf, Bruno
and Rybarczyk, Th{\'e}o
and Gleyzes, S{\'e}bastien
and Rouchon, Pierre
and Mirrahimi, Mazyar
and Amini, Hadis
and Brune, Michel
and Raimond, Jean-Michel
and Haroche, Serge},
title={Real-time quantum feedback prepares and stabilizes photon number states},
journal={Nature},
year={2011},
month={09},
day={01},
volume={477},
number={7362},
pages={73-77},
issn={1476-4687},
doi={10.1038/nature10376},
url={https://doi.org/10.1038/nature10376}
}

@article{Zhou2012PRL,
  title = {Field Locked to a Fock State by Quantum Feedback with Single Photon Corrections},
  author = {Zhou, X. and Dotsenko, I. and Peaudecerf, B. and Rybarczyk, T. and Sayrin, C. and Gleyzes, S. and Raimond, J. M. and Brune, M. and Haroche, S.},
  journal = {Phys. Rev. Lett.},
  volume = {108},
  issue = {24},
  pages = {243602},
  numpages = {5},
  year = {2012},
  month = {06},
  publisher = {American Physical Society},
  doi = {10.1103/PhysRevLett.108.243602},
  url = {https://link.aps.org/doi/10.1103/PhysRevLett.108.243602}
}

@article{Inoue2013PRL,
  title = {Unconditional Quantum-Noise Suppression via Measurement-Based Quantum Feedback},
  author = {Inoue, Ryotaro and Tanaka, Shin-Ichi-Ro and Namiki, Ryo and Sagawa, Takahiro and Takahashi, Yoshiro},
  journal = {Phys. Rev. Lett.},
  volume = {110},
  issue = {16},
  pages = {163602},
  numpages = {5},
  year = {2013},
  month = {04},
  publisher = {American Physical Society},
  doi = {10.1103/PhysRevLett.110.163602},
  url = {https://link.aps.org/doi/10.1103/PhysRevLett.110.163602}
}

@Article{Riste2013nature,
author={Rist{\`e}, D.
and Dukalski, M.
and Watson, C. A.
and de Lange, G.
and Tiggelman, M. J.
and Blanter, Ya. M.
and Lehnert, K. W.
and Schouten, R. N.
and DiCarlo, L.},
title={Deterministic entanglement of superconducting qubits by parity measurement and feedback},
journal={Nature},
year={2013},
month={10},
day={01},
volume={502},
number={7471},
pages={350-354},
issn={1476-4687},
doi={10.1038/nature12513},
url={https://doi.org/10.1038/nature12513}
}

@article{Wade2015PRL,
  title = {Squeezing and Entanglement of Density Oscillations in a Bose-Einstein Condensate},
  author = {Wade, Andrew C. J. and Sherson, Jacob F. and M\o{}lmer, Klaus},
  journal = {Phys. Rev. Lett.},
  volume = {115},
  issue = {6},
  pages = {060401},
  numpages = {6},
  year = {2015},
  month = {08},
  publisher = {American Physical Society},
  doi = {10.1103/PhysRevLett.115.060401},
  url = {https://link.aps.org/doi/10.1103/PhysRevLett.115.060401}
}

@article{Cox2016PRL,
  title = {Deterministic Squeezed States with Collective Measurements and Feedback},
  author = {Cox, Kevin C. and Greve, Graham P. and Weiner, Joshua M. and Thompson, James K.},
  journal = {Phys. Rev. Lett.},
  volume = {116},
  issue = {9},
  pages = {093602},
  numpages = {5},
  year = {2016},
  month = {03},
  publisher = {American Physical Society},
  doi = {10.1103/PhysRevLett.116.093602},
  url = {https://link.aps.org/doi/10.1103/PhysRevLett.116.093602}
}

@article{Gajdacz2016PRL,
  title = {Preparation of Ultracold Atom Clouds at the Shot Noise Level},
  author = {Gajdacz, M. and Hilliard, A. J. and Kristensen, M. A. and Pedersen, P. L. and Klempt, C. and Arlt, J. J. and Sherson, J. F.},
  journal = {Phys. Rev. Lett.},
  volume = {117},
  issue = {7},
  pages = {073604},
  numpages = {5},
  year = {2016},
  month = {08},
  publisher = {American Physical Society},
  doi = {10.1103/PhysRevLett.117.073604},
  url = {https://link.aps.org/doi/10.1103/PhysRevLett.117.073604}
}

@article{Lammers2016PRA,
  title = {Open-system many-body dynamics through interferometric measurements and feedback},
  author = {Lammers, Jonas and Weimer, Hendrik and Hammerer, Klemens},
  journal = {Phys. Rev. A},
  volume = {94},
  issue = {5},
  pages = {052120},
  numpages = {16},
  year = {2016},
  month = {11},
  publisher = {American Physical Society},
  doi = {10.1103/PhysRevA.94.052120},
  url = {https://link.aps.org/doi/10.1103/PhysRevA.94.052120}
}

@article{Sudhir2017PRX,
  title = {Appearance and Disappearance of Quantum Correlations in Measurement-Based Feedback Control of a Mechanical Oscillator},
  author = {Sudhir, V. and Wilson, D. J. and Schilling, R. and Sch\"utz, H. and Fedorov, S. A. and Ghadimi, A. H. and Nunnenkamp, A. and Kippenberg, T. J.},
  journal = {Phys. Rev. X},
  volume = {7},
  issue = {1},
  pages = {011001},
  numpages = {14},
  year = {2017},
  month = {01},
  publisher = {American Physical Society},
  doi = {10.1103/PhysRevX.7.011001},
  url = {https://link.aps.org/doi/10.1103/PhysRevX.7.011001}
}

@article{Borah2021PRL,
  title = {Measurement-Based Feedback Quantum Control with Deep Reinforcement Learning for a Double-Well Nonlinear Potential},
  author = {Borah, Sangkha and Sarma, Bijita and Kewming, Michael and Milburn, Gerard J. and Twamley, Jason},
  journal = {Phys. Rev. Lett.},
  volume = {127},
  issue = {19},
  pages = {190403},
  numpages = {6},
  year = {2021},
  month = {11},
  publisher = {American Physical Society},
  doi = {10.1103/PhysRevLett.127.190403},
  url = {https://link.aps.org/doi/10.1103/PhysRevLett.127.190403}
}

@article{Porotti2022quantum,
  doi = {10.22331/q-2022-06-28-747},
  url = {https://doi.org/10.22331/q-2022-06-28-747},
  title = {Deep {R}einforcement {L}earning for {Q}uantum {S}tate {P}reparation with {W}eak {N}onlinear {M}easurements},
  author = {Porotti, Riccardo and Essig, Antoine and Huard, Benjamin and Marquardt, Florian},
  journal = {{Quantum}},
  issn = {2521-327X},
  publisher = {{Verein zur F{\"{o}}rderung des Open Access Publizierens in den Quantenwissenschaften}},
  volume = {6},
  pages = {747},
  month = jun,
  year = {2022}
}

@article{Fosel2018PRX,
  title = {Reinforcement Learning with Neural Networks for Quantum Feedback},
  author = {F\"osel, Thomas and Tighineanu, Petru and Weiss, Talitha and Marquardt, Florian},
  journal = {Phys. Rev. X},
  volume = {8},
  issue = {3},
  pages = {031084},
  numpages = {15},
  year = {2018},
  month = {09},
  publisher = {American Physical Society},
  doi = {10.1103/PhysRevX.8.031084},
  url = {https://link.aps.org/doi/10.1103/PhysRevX.8.031084}
}

@article{Hutin2025PRXQuantum,
  title = {Preparing Schr\"odinger Cat States in a Microwave Cavity Using a Neural Network},
  author = {Hutin, Hector and Bilous, Pavlo and Ye, Chengzhi and Abdollahi, Sepideh and Cros, Loris and Dvir, Tom and Shah, Tirth and Cohen, Yonatan and Bienfait, Audrey and Marquardt, Florian and Huard, Benjamin},
  journal = {PRX Quantum},
  volume = {6},
  issue = {1},
  pages = {010321},
  numpages = {22},
  year = {2025},
  month = {01},
  publisher = {American Physical Society},
  doi = {10.1103/PRXQuantum.6.010321},
  url = {https://link.aps.org/doi/10.1103/PRXQuantum.6.010321}
}

@article{PhysRevA.95.043641,
  title = {Monte Carlo wave-function description of losses in a one-dimensional Bose gas and cooling to the ground state by quantum feedback},
  author = {Schemmer, M. and Johnson, A. and Photopoulos, R. and Bouchoule, I.},
  journal = {Phys. Rev. A},
  volume = {95},
  issue = {4},
  pages = {043641},
  numpages = {7},
  year = {2017},
  month = {04},
  publisher = {American Physical Society},
  doi = {10.1103/PhysRevA.95.043641},
  url = {https://link.aps.org/doi/10.1103/PhysRevA.95.043641}
}

@article{Hush_2013,
	doi = {10.1088/1367-2630/15/11/113060},
	url = {https://doi.org/10.1088/1367-2630/15/11/113060},
	year = 2013,
	month = {11},
	publisher = {{IOP} Publishing},
	volume = {15},
	number = {11},
	pages = {113060},
	author = {M R Hush and S S Szigeti and A R R Carvalho and J J Hope},
	title = {Controlling spontaneous-emission noise in measurement-based feedback cooling of a Bose{\textendash}Einstein condensate},
	journal = {New Journal of Physics}
}

@article{PhysRevA.75.051405,
  title = {External-feedback laser cooling of molecular gases},
  author = {Vuleti\ifmmode \acute{c}\else \'{c}\fi{}, Vladan and Thompson, James K. and Black, Adam T. and Simon, Jonathan},
  journal = {Phys. Rev. A},
  volume = {75},
  issue = {5},
  pages = {051405},
  numpages = {4},
  year = {2007},
  month = {05},
  publisher = {American Physical Society},
  doi = {10.1103/PhysRevA.75.051405},
  url = {https://link.aps.org/doi/10.1103/PhysRevA.75.051405}
}

@article{PhysRevA.100.063819,
  title = {Measurement and feedback for cooling heavy levitated particles in low-frequency traps},
  author = {Walker, L. S. and Robb, G. R. M. and Daley, A. J.},
  journal = {Phys. Rev. A},
  volume = {100},
  issue = {6},
  pages = {063819},
  numpages = {9},
  year = {2019},
  month = {12},
  publisher = {American Physical Society},
  doi = {10.1103/PhysRevA.100.063819},
  url = {https://link.aps.org/doi/10.1103/PhysRevA.100.063819}
}

@article{PhysRevLett.96.043003,
  title = {Feedback Cooling of a Single Trapped Ion},
  author = {Bushev, Pavel and Rotter, Daniel and Wilson, Alex and Dubin, Fran\ifmmode \mbox{\c{c}}\else \c{c}\fi{}ois and Becher, Christoph and Eschner, J\"urgen and Blatt, Rainer and Steixner, Viktor and Rabl, Peter and Zoller, Peter},
  journal = {Phys. Rev. Lett.},
  volume = {96},
  issue = {4},
  pages = {043003},
  numpages = {4},
  year = {2006},
  month = {02},
  publisher = {American Physical Society},
  doi = {10.1103/PhysRevLett.96.043003},
  url = {https://link.aps.org/doi/10.1103/PhysRevLett.96.043003}
}

@article{PhysRevLett.105.173003,
  title = {Feedback Cooling of a Single Neutral Atom},
  author = {Koch, Markus and Sames, Christian and Kubanek, Alexander and Apel, Matthias and Balbach, Maximilian and Ourjoumtsev, Alexei and Pinkse, Pepijn W. H. and Rempe, Gerhard},
  journal = {Phys. Rev. Lett.},
  volume = {105},
  issue = {17},
  pages = {173003},
  numpages = {4},
  year = {2010},
  month = {10},
  publisher = {American Physical Society},
  doi = {10.1103/PhysRevLett.105.173003},
  url = {https://link.aps.org/doi/10.1103/PhysRevLett.105.173003}
}

@Article{Wilson2015,
author={Wilson, D. J.
and Sudhir, V.
and Piro, N.
and Schilling, R.
and Ghadimi, A.
and Kippenberg, T. J.},
title={Measurement-based control of a mechanical oscillator at its thermal decoherence rate},
journal={Nature},
year={2015},
month={08},
day={01},
volume={524},
number={7565},
pages={325-329},
issn={1476-4687},
doi={10.1038/nature14672},
url={https://doi.org/10.1038/nature14672}
}

@article{PhysRevLett.124.110503,
  title = {Simulating Nonlinear Dynamics of Collective Spins via Quantum Measurement and Feedback},
  author = {Mu\~noz-Arias, Manuel H. and Poggi, Pablo M. and Jessen, Poul S. and Deutsch, Ivan H.},
  journal = {Phys. Rev. Lett.},
  volume = {124},
  issue = {11},
  pages = {110503},
  numpages = {7},
  year = {2020},
  month = {03},
  publisher = {American Physical Society},
  doi = {10.1103/PhysRevLett.124.110503},
  url = {https://link.aps.org/doi/10.1103/PhysRevLett.124.110503}
}

@article{PhysRevA.102.022610,
  title = {Simulation of the complex dynamics of mean-field $p$-spin models using measurement-based quantum feedback control},
  author = {Mu\~noz-Arias, Manuel H. and Deutsch, Ivan H. and Jessen, Poul S. and Poggi, Pablo M.},
  journal = {Phys. Rev. A},
  volume = {102},
  issue = {2},
  pages = {022610},
  numpages = {15},
  year = {2020},
  month = {08},
  publisher = {American Physical Society},
  doi = {10.1103/PhysRevA.102.022610},
  url = {https://link.aps.org/doi/10.1103/PhysRevA.102.022610}
}

@Article{Vijay2012,
author={Vijay, R.
and Macklin, C.
and Slichter, D. H.
and Weber, S. J.
and Murch, K. W.
and Naik, R.
and Korotkov, A. N.
and Siddiqi, I.},
title={Stabilizing Rabi oscillations in a superconducting qubit using quantum feedback},
journal={Nature},
year={2012},
month={10},
day={01},
volume={490},
number={7418},
pages={77-80},
issn={1476-4687},
doi={10.1038/nature11505},
url={https://doi.org/10.1038/nature11505}
}

@article{PhysRevLett.88.093003,
  title = {Feedback Control of Atomic Motion in an Optical Lattice},
  author = {Morrow, N. V. and Dutta, S. K. and Raithel, G.},
  journal = {Phys. Rev. Lett.},
  volume = {88},
  issue = {9},
  pages = {093003},
  numpages = {4},
  year = {2002},
  month = {02},
  publisher = {American Physical Society},
  doi = {10.1103/PhysRevLett.88.093003},
  url = {https://link.aps.org/doi/10.1103/PhysRevLett.88.093003}
}

@article{PhysRevLett.110.210503,
  title = {Feedback Control of Trapped Coherent Atomic Ensembles},
  author = {Vanderbruggen, T. and Kohlhaas, R. and Bertoldi, A. and Bernon, S. and Aspect, A. and Landragin, A. and Bouyer, P.},
  journal = {Phys. Rev. Lett.},
  volume = {110},
  issue = {21},
  pages = {210503},
  numpages = {5},
  year = {2013},
  month = {05},
  publisher = {American Physical Society},
  doi = {10.1103/PhysRevLett.110.210503},
  url = {https://link.aps.org/doi/10.1103/PhysRevLett.110.210503}
}

@article{Kroeger_2020,
	doi = {10.1088/1367-2630/ab73cc},
	url = {https://doi.org/10.1088%2F1367-2630%2Fab73cc},
	year = 2020,
	month = {03},
	publisher = {{IOP} Publishing},
	volume = {22},
	number = {3},
	pages = {033020},
	author = {Katrin Kroeger and Nishant Dogra and Rodrigo Rosa-Medina and Marcin Paluch and Francesco Ferri and Tobias Donner and Tilman Esslinger},
	title = {Continuous feedback on a quantum gas coupled to an optical cavity},
	journal = {New Journal of Physics},
}

@article{Kopylov_2015,
	doi = {10.1088/1367-2630/17/1/013040},
	url = {https://doi.org/10.1088/1367-2630/17/1/013040},
	year = 2015,
	month = {01},
	publisher = {{IOP} Publishing},
	volume = {17},
	number = {1},
	pages = {013040},
	author = {Wassilij Kopylov and Clive Emary and Eckehard Schöll and Tobias Brandes},
	title = {Time-delayed feedback control of the Dicke{\textendash}Hepp{\textendash}Lieb superradiant quantum phase transition},
	journal = {New Journal of Physics}
}

@article{PhysRevLett.124.010603,
  title = {Feedback-Induced Quantum Phase Transitions Using Weak Measurements},
  author = {Ivanov, D. A. and Ivanova, T. Yu. and Caballero-Benitez, S. F. and Mekhov, I. B.},
  journal = {Phys. Rev. Lett.},
  volume = {124},
  issue = {1},
  pages = {010603},
  numpages = {7},
  year = {2020},
  month = {01},
  publisher = {American Physical Society},
  doi = {10.1103/PhysRevLett.124.010603},
  url = {https://link.aps.org/doi/10.1103/PhysRevLett.124.010603}
}

@article{PhysRevResearch.2.043325,
  title = {Feedback induced magnetic phases in binary Bose-Einstein condensates},
  author = {Hurst, Hilary M. and Guo, Shangjie and Spielman, I. B.},
  journal = {Phys. Rev. Research},
  volume = {2},
  issue = {4},
  pages = {043325},
  numpages = {12},
  year = {2020},
  month = {12},
  publisher = {American Physical Society},
  doi = {10.1103/PhysRevResearch.2.043325},
  url = {https://link.aps.org/doi/10.1103/PhysRevResearch.2.043325}
}

@article{PhysRevA.99.053612,
  title = {Measurement-induced dynamics and stabilization of spinor-condensate domain walls},
  author = {Hurst, Hilary M. and Spielman, I. B.},
  journal = {Phys. Rev. A},
  volume = {99},
  issue = {5},
  pages = {053612},
  numpages = {6},
  year = {2019},
  month = {05},
  publisher = {American Physical Society},
  doi = {10.1103/PhysRevA.99.053612},
  url = {https://link.aps.org/doi/10.1103/PhysRevA.99.053612}
}

@article{Young2021PRR,
  title = {Feedback-stabilized dynamical steady states in the Bose-Hubbard model},
  author = {Young, Jeremy T. and Gorshkov, Alexey V. and Spielman, I. B.},
  journal = {Phys. Rev. Res.},
  volume = {3},
  issue = {4},
  pages = {043075},
  numpages = {13},
  year = {2021},
  month = {10},
  publisher = {American Physical Society},
  doi = {10.1103/PhysRevResearch.3.043075},
  url = {https://link.aps.org/doi/10.1103/PhysRevResearch.3.043075}
}

@article{PhysRevA.76.010301,
  title = {Stabilizing entanglement by quantum-jump-based feedback},
  author = {Carvalho, A. R. R. and Hope, J. J.},
  journal = {Phys. Rev. A},
  volume = {76},
  issue = {1},
  pages = {010301},
  numpages = {4},
  year = {2007},
  month = {07},
  publisher = {American Physical Society},
  doi = {10.1103/PhysRevA.76.010301},
  url = {https://link.aps.org/doi/10.1103/PhysRevA.76.010301}
}

@Article{Wang2010,
author={Wang, L. C.
and Shen, J.
and Yi, X. X.},
title={Effect of feedback control on the entanglement evolution},
journal={The European Physical Journal D},
year={2010},
month={02},
day={01},
volume={56},
number={3},
pages={435-440},
issn={1434-6079},
doi={10.1140/epjd/e2009-00307-2},
url={https://doi.org/10.1140/epjd/e2009-00307-2}
}

@article{Sivak2022PRX,
  title = {Model-Free Quantum Control with Reinforcement Learning},
  author = {Sivak, V. V. and Eickbusch, A. and Liu, H. and Royer, B. and Tsioutsios, I. and Devoret, M. H.},
  journal = {Phys. Rev. X},
  volume = {12},
  issue = {1},
  pages = {011059},
  numpages = {23},
  year = {2022},
  month = {03},
  publisher = {American Physical Society},
  doi = {10.1103/PhysRevX.12.011059},
  url = {https://link.aps.org/doi/10.1103/PhysRevX.12.011059}
}

@Article{Reuer2023NC,
author={Reuer, Kevin
and Landgraf, Jonas
and F{\"o}sel, Thomas
and O'Sullivan, James
and Beltr{\'a}n, Liberto
and Akin, Abdulkadir
and Norris, Graham J.
and Remm, Ants
and Kerschbaum, Michael
and Besse, Jean-Claude
and Marquardt, Florian
and Wallraff, Andreas
and Eichler, Christopher},
title={Realizing a deep reinforcement learning agent for real-time quantum feedback},
journal={Nature Communications},
year={2023},
month={11},
day={06},
volume={14},
number={1},
pages={7138},
issn={2041-1723},
doi={10.1038/s41467-023-42901-3},
url={https://doi.org/10.1038/s41467-023-42901-3}
}

@article{PhysRevA.78.012334,
  title = {Controlling entanglement by direct quantum feedback},
  author = {Carvalho, A. R. R. and Reid, A. J. S. and Hope, J. J.},
  journal = {Phys. Rev. A},
  volume = {78},
  issue = {1},
  pages = {012334},
  numpages = {10},
  year = {2008},
  month = {07},
  publisher = {American Physical Society},
  doi = {10.1103/PhysRevA.78.012334},
  url = {https://link.aps.org/doi/10.1103/PhysRevA.78.012334}
}

@article{Yamaguchi2023PRA,
  title = {Feedback-cooled Bose-Einstein condensation: Near and far from equilibrium},
  author = {Yamaguchi, Evan P. and Hurst, Hilary M. and Spielman, I. B.},
  journal = {Phys. Rev. A},
  volume = {107},
  issue = {6},
  pages = {063306},
  numpages = {13},
  year = {2023},
  month = {06},
  publisher = {American Physical Society},
  doi = {10.1103/PhysRevA.107.063306},
  url = {https://link.aps.org/doi/10.1103/PhysRevA.107.063306}
}

@article{Buonaiuto2021PRL,
  title = {Dynamical Phases and Quantum Correlations in an Emitter-Waveguide System with Feedback},
  author = {Buonaiuto, Giuseppe and Carollo, Federico and Olmos, Beatriz and Lesanovsky, Igor},
  journal = {Phys. Rev. Lett.},
  volume = {127},
  issue = {13},
  pages = {133601},
  numpages = {8},
  year = {2021},
  month = {09},
  publisher = {American Physical Society},
  doi = {10.1103/PhysRevLett.127.133601},
  url = {https://link.aps.org/doi/10.1103/PhysRevLett.127.133601}
}

@article{PhysRevLett.92.223004,
  title = {Quantum Feedback Control of Atomic Motion in an Optical Cavity},
  author = {Steck, Daniel A. and Jacobs, Kurt and Mabuchi, Hideo and Bhattacharya, Tanmoy and Habib, Salman},
  journal = {Phys. Rev. Lett.},
  volume = {92},
  issue = {22},
  pages = {223004},
  numpages = {4},
  year = {2004},
  month = {06},
  publisher = {American Physical Society},
  doi = {10.1103/PhysRevLett.92.223004},
  url = {https://link.aps.org/doi/10.1103/PhysRevLett.92.223004}
}

@article{PhysRevLett.110.013601,
  title = {Single Photon Delayed Feedback: A Way to Stabilize Intrinsic Quantum Cavity Electrodynamics},
  author = {Carmele, Alexander and Kabuss, Julia and Schulze, Franz and Reitzenstein, Stephan and Knorr, Andreas},
  journal = {Phys. Rev. Lett.},
  volume = {110},
  issue = {1},
  pages = {013601},
  numpages = {5},
  year = {2013},
  month = {01},
  publisher = {American Physical Society},
  doi = {10.1103/PhysRevLett.110.013601},
  url = {https://link.aps.org/doi/10.1103/PhysRevLett.110.013601}
}

@article{PhysRevA.62.012307,
  title = {Quantum feedback with weak measurements},
  author = {Lloyd, Seth and Slotine, Jean-Jacques E.},
  journal = {Phys. Rev. A},
  volume = {62},
  issue = {1},
  pages = {012307},
  numpages = {5},
  year = {2000},
  month = {06},
  publisher = {American Physical Society},
  doi = {10.1103/PhysRevA.62.012307},
  url = {https://link.aps.org/doi/10.1103/PhysRevA.62.012307}
}

@article{PhysRevLett.111.103601,
  title = {Feedback Cooling of an Atomic Spin Ensemble},
  author = {Behbood, N. and Colangelo, G. and Martin Ciurana, F. and Napolitano, M. and Sewell, R. J. and Mitchell, M. W.},
  journal = {Phys. Rev. Lett.},
  volume = {111},
  issue = {10},
  pages = {103601},
  numpages = {5},
  year = {2013},
  month = {09},
  publisher = {American Physical Society},
  doi = {10.1103/PhysRevLett.111.103601},
  url = {https://link.aps.org/doi/10.1103/PhysRevLett.111.103601}
}

@article{Ivanov_2014,
	doi = {10.1088/0953-4075/47/13/135303},
	url = {https://doi.org/10.1088/0953-4075/47/13/135303},
	year = 2014,
	month = {06},
	publisher = {{IOP} Publishing},
	volume = {47},
	number = {13},
	pages = {135303},
	author = {D A Ivanov and T Yu Ivanova},
	title = {Bragg-reflection-based feedback cooling of optically trapped particles},
	journal = {Journal of Physics B: Atomic, Molecular and Optical Physics}
}

@article{PhysRevA.80.013614,
  title = {Continuous measurement feedback control of a Bose-Einstein condensate using phase-contrast imaging},
  author = {Szigeti, S. S. and Hush, M. R. and Carvalho, A. R. R. and Hope, J. J.},
  journal = {Phys. Rev. A},
  volume = {80},
  issue = {1},
  pages = {013614},
  numpages = {11},
  year = {2009},
  month = {07},
  publisher = {American Physical Society},
  doi = {10.1103/PhysRevA.80.013614},
  url = {https://link.aps.org/doi/10.1103/PhysRevA.80.013614}
}

@article{vorberg,
  title = {Nonequilibrium steady states of ideal bosonic and fermionic quantum gases},
  author = {Vorberg, D. and Wustmann, W. and Schomerus, H. and Ketzmerick, R. and Eckardt, A.},
  journal = {Phys. Rev. E},
  volume = {92},
  pages = {062119},
  year = {2015},
  publisher = {American Physical Society},
  doi = {10.1103/PhysRevE.92.062119},
  url = {https://link.aps.org/doi/10.1103/PhysRevE.92.062119}
}

@article{vonKlitzing1980,
  title = {New Method for High-Accuracy Determination of the Fine-Structure Constant Based on Quantized Hall Resistance},
  author = {Klitzing, K. v. and Dorda, G. and Pepper, M.},
  journal = {Phys. Rev. Lett.},
  volume = {45},
  issue = {6},
  pages = {494},
  numpages = {0},
  year = {1980},
  publisher = {American Physical Society},
  doi = {10.1103/PhysRevLett.45.494},
  url = {https://link.aps.org/doi/10.1103/PhysRevLett.45.494}
}

@article{Seetharam2015,
  title = {Controlled Population of Floquet-Bloch States via Coupling to Bose and Fermi Baths},
  author = {Seetharam, K.~I. and Bardyn, C.-E. and Lindner, Netanel H. and Rudner, Mark S. and Refael, Gil},
  journal = {Phys. Rev. X},
  volume = {5},
  issue = {4},
  pages = {041050},
  numpages = {27},
  year = {2015},
  publisher = {American Physical Society},
  doi = {10.1103/PhysRevX.5.041050},
  url = {https://link.aps.org/doi/10.1103/PhysRevX.5.041050}
}

@article{Tai2017,
	title = {Microscopy of the interacting {Harper}–{Hofstadter} model in the two-body limit},
	volume = {546},
	issn = {1476-4687},
	url = {https://www.nature.com/articles/nature22811},
	doi = {10.1038/nature22811},
	journal = {Nature},
	author = {Tai, M. Eric and Lukin, Alexander and Rispoli, Matthew and Schittko, Robert and Menke, Tim and {Dan Borgnia} and Preiss, Philipp M. and Grusdt, Fabian and Kaufman, Adam M. and Greiner, Markus},
	year = {2017},
	pages = {519--523},
}

@article{Wu2022h,
doi = {10.1088/1367-2630/aca81e},
url = {https://dx.doi.org/10.1088/1367-2630/aca81e},
year = {2022},
publisher = {IOP Publishing},
volume = {24},
pages = {123015},
author = {L.-N. Wu and A. Eckardt},
title = {Heat transport in an optical lattice via Markovian feedback control},
journal = {New J. Phys.}
}

@Article{Schnell2024,
	title={{Dissipative preparation of a Floquet topological insulator in an optical lattice via bath engineering}},
	author={A. Schnell and C. Weitenberg and A. Eckardt},
	journal={SciPost Phys.},
	volume={17},
	pages={052},
	year={2024},
	publisher={SciPost},
	doi={10.21468/SciPostPhys.17.2.052},
	url={https://scipost.org/10.21468/SciPostPhys.17.2.052},
}

@article{Qin2018,
  title = {Charge density wave and charge pump of interacting fermions in circularly shaken hexagonal optical lattices},
  author = {Qin, T. and Schnell, A. and Sengstock, K. and Weitenberg, C. and Eckardt, A. and Hofstetter, W.},
  journal = {Phys. Rev. A},
  volume = {98},
  issue = {3},
  pages = {033601},
  numpages = {11},
  year = {2018},
  publisher = {American Physical Society},
  doi = {10.1103/PhysRevA.98.033601},
  url = {https://link.aps.org/doi/10.1103/PhysRevA.98.033601}
}

@article{Schnell2023,
  title = {Floquet-heating-induced Bose condensation in a scarlike mode of an open driven optical-lattice system},
  author = {Schnell, A. and Wu, L.-N. and Widera, A. and Eckardt, A.},
  journal = {Phys. Rev. A},
  volume = {107},
  issue = {2},
  pages = {L021301},
  numpages = {6},
  year = {2023},
  publisher = {American Physical Society},
  doi = {10.1103/PhysRevA.107.L021301},
  url = {https://link.aps.org/doi/10.1103/PhysRevA.107.L021301}
}

@article{Petiziol2024c,
  title = {Controlling nonequilibrium Bose-Einstein condensation with engineered environments},
  author = {Petiziol, F. and Eckardt, A.},
  journal = {Phys. Rev. A},
  volume = {110},
  issue = {2},
  pages = {L021701},
  numpages = {7},
  year = {2024},
  publisher = {American Physical Society},
  doi = {10.1103/PhysRevA.110.L021701},
  url = {https://link.aps.org/doi/10.1103/PhysRevA.110.L021701}
}

@article{Petiziol2022,
  title = {Cavity-Based Reservoir Engineering for Floquet-Engineered Superconducting Circuits},
  author = {Petiziol, F. and Eckardt, A.},
  journal = {Phys. Rev. Lett.},
  volume = {129},
  issue = {23},
  pages = {233601},
  numpages = {7},
  year = {2022},
  publisher = {American Physical Society},
  doi = {10.1103/PhysRevLett.129.233601},
  url ={https://link.aps.org/doi/10.1103/PhysRevLett.129.233601}
}

@article{Kapit2014,
  title = {Induced Self-Stabilization in Fractional Quantum Hall States of Light},
  author = {Kapit, E. and Hafezi, M. and Simon, S.~H.},
  journal = {Phys. Rev. X},
  volume = {4},
  issue = {3},
  pages = {031039},
  numpages = {11},
  year = {2014},
  publisher = {American Physical Society},
  doi = {10.1103/PhysRevX.4.031039},
  url = {https://link.aps.org/doi/10.1103/PhysRevX.4.031039}
}

@article{Kraus2008,
  title = {Preparation of entangled states by quantum Markov processes},
  author = {Kraus, B. and B\"uchler, H. P. and Diehl, S. and Kantian, A. and Micheli, A. and Zoller, P.},
  journal = {Phys. Rev. A},
  volume = {78},
  issue = {4},
  pages = {042307},
  numpages = {9},
  year = {2008},
  publisher = {American Physical Society},
  doi = {10.1103/PhysRevA.78.042307},
  url = {https://link.aps.org/doi/10.1103/PhysRevA.78.042307}
}

@article{RevModPhys.89.011004,
  title = {Colloquium: Atomic quantum gases in periodically driven optical lattices},
  author = {Eckardt, A.},
  journal = {Rev. Mod. Phys.},
  volume = {89},
  issue = {1},
  pages = {011004},
  numpages = {30},
  year = {2017},
  publisher = {American Physical Society},
  doi = {10.1103/RevModPhys.89.011004},
  url = {https://link.aps.org/doi/10.1103/RevModPhys.89.011004}
}

@article{Aidelsburger2015,
  title = {{Measuring the Chern number of Hofstadter bands with ultracold bosonic atoms}},
  author = {Aidelsburger, M. and Lohse, M. and Schweizer, C. and Atala, M. and Barreiro, J. T. and Nascimbène, S. and Cooper, N. R. and Bloch, I. and Goldman, N.},
  journal = {Nat. Phys.},
  volume = {11},
  pages = {162-166},
  year = {2015},
  doi = {10.1038/nphys3171},
  url = {https://doi.org/10.1038/nphys3171}
}

@article{Jotzu2014,
author = {Jotzu, G. and Messer, Michael and Desbuquois, R{\'{e}}mi and Lebrat, Martin and Uehlinger, Thomas and Greif, Daniel and Esslinger, Tilman},
doi = {10.1038/nature13915},
url = {https://doi.org/10.1038/nature13915},
issn = {14764687},
journal = {Nature},
pages = {237--240},
title = {{Experimental realization of the topological Haldane model with ultracold fermions}},
volume = {515},
year = {2014}
}

@article{Wintersperger2020,
  title = {{Realization of an anomalous Floquet topological system with ultracold atoms}},
  author = {Wintersperger, K. and Braun, C. and {\"U}nal, F. N. and Eckardt, A. and Di Liberto, M. and Goldman, N. and Bloch, I. and Aidelsburger, M.},
  journal = {Nat. Phys.},
  volume = {16},
  pages = {1058–1063},
  year = {2020},
  doi = {10.1038/s41567-020-0949-y},
  url = {https://www.nature.com/articles/s41567-020-0949-y#citeas}
}

@article{lnwu,
  title = {Cooling and state preparation in an optical lattice via Markovian feedback control},
  author = {Wu, L.-N. and Eckardt, A.},
  journal = {Phys. Rev. Res.},
  volume = {4},
  issue = {2},
  pages = {L022045},
  numpages = {7},
  year = {2022},
  publisher = {American Physical Society},
  doi = {10.1103/PhysRevResearch.4.L022045},
  url = {https://link.aps.org/doi/10.1103/PhysRevResearch.4.L022045}
}

@article{RiceMele,
  title = {Elementary Excitations of a Linearly Conjugated Diatomic Polymer},
  author = {Rice, M. J. and Mele, E. J.},
  journal = {Phys. Rev. Lett.},
  volume = {49},
  issue = {19},
  pages = {1455},
  numpages = {0},
  year = {1982},
  month= {11},
  publisher = {American Physical Society},
  doi = {10.1103/PhysRevLett.49.1455},
  url = {https://link.aps.org/doi/10.1103/PhysRevLett.49.1455}
}

@article{Hasan2010,
  title = {Colloquium: Topological insulators},
  author = {Hasan, M. Z. and Kane, C. L.},
  journal = {Rev. Mod. Phys.},
  volume = {82},
  issue = {4},
  pages = {3045},
  numpages = {0},
  year = {2010},
  publisher = {American Physical Society},
  doi = {10.1103/RevModPhys.82.3045},
  url = {https://link.aps.org/doi/10.1103/RevModPhys.82.3045}
}

@article{PhysRevLett.111.240405,
  title = {Generalized Bose-Einstein Condensation into Multiple States in Driven-Dissipative Systems},
  author = {Vorberg, D. and Wustmann, W. and Ketzmerick, R. and Eckardt, A.},
  journal = {Phys. Rev. Lett.},
  volume = {111},
  issue = {24},
  pages = {240405},
  numpages = {5},
  year = {2013},
  month= {12},
  publisher = {American Physical Society},
  doi = {10.1103/PhysRevLett.111.240405},
  url = {https://link.aps.org/doi/10.1103/PhysRevLett.111.240405}
}

@book{PrangeGirvin1990, 
place={Berlin}, 
title={The Quantum Hall Effect}, 
doi={10.1007/978-1-4612-3350-3}, 
url = { https://link.springer.com/book/10.1007/978-1-4612-3350-3},
publisher={Springer-Verlag}, 
author={Prange, R.~E. and Girvin, S.~M.}, 
year={1990}}

@book{wiseman_milburn_2009, 
place={Cambridge}, 
title={Quantum Measurement and Control}, 
doi={10.1017/CBO9780511813948}, 
url = { https://doi.org/10.1017/CBO9780511813948},
publisher={Cambridge University Press}, 
author={Wiseman, H.~M. and Milburn, G.~J.}, 
year={2009}}

@article{Wiseman,
  title = {Quantum theory of continuous feedback},
  author = {Wiseman, H. M.},
  journal = {Phys. Rev. A},
  volume = {49},
  issue = {3},
  pages = {2133},
  numpages = {0},
  year = {1994},
  month= {},
  publisher = {American Physical Society},
  doi = {10.1103/PhysRevA.49.2133},
  url = {https://link.aps.org/doi/10.1103/PhysRevA.49.2133}
}

@book{Asb_th_2016,
	doi = {10.1007/978-3-319-25607-8},  
	url = {https://doi.org/10.1007%2F978-3-319-25607-8},  
	year = 2016,
	publisher = {Springer International Publishing},  
	author = {J.~K. Asb{\'{o}}th and L. Oroszl{\'{a}}ny and A. P{\'{a}}lyi},  
	title = {A Short Course on Topological Insulators}
}

@article{Kohn,
  title = {Analytic Properties of Bloch Waves and Wannier Functions},
  author = {Kohn, W.},
  journal = {Phys. Rev.},
  volume = {115},
  issue = {4},
  pages = {809},
  year = {1959},
  publisher = {American Physical Society},
  doi = {10.1103/PhysRev.115.809},
  url = {https://link.aps.org/doi/10.1103/PhysRev.115.809}
}

@article{Modugno,
	doi = {10.1088/1367-2630/14/5/055004},
	url = {https://doi.org/10.1088/1367-2630/14/5/055004},
	year = 2012,
	publisher = {{IOP} Publishing},
	volume = {14},
	pages = {055004},
	author = {M. Modugno and G. Pettini},
	title = {Maximally localized Wannier functions for ultracold atoms in one-dimensional double-well periodic potentials},
	journal = {New J. Phys.}
	}

@article{PhysRevA.72.024104,
  title = {Parametrization of the feedback Hamiltonian realizing a pure steady state},
  author = {Yamamoto, N.},
  journal = {Phys. Rev. A},
  volume = {72},
  pages = {024104},
  numpages = {4},
  year = {2005},
  publisher = {American Physical Society},
  doi = {10.1103/PhysRevA.72.024104},
  url = {https://link.aps.org/doi/10.1103/PhysRevA.72.024104}
}

@article{Bloch2012,
author={Bloch, I.
and Dalibard, J.
and Nascimb{\`e}ne, S.},
title={Quantum simulations with ultracold quantum gases},
journal={Nat. Phys.},
year={2012},
day={01},
volume={8},
pages={267-276},
issn={1745-2481},
doi={10.1038/nphys2259},
url={https://doi.org/10.1038/nphys2259}
}

@article{PhysRevA.47.642,
  title = {Quantum theory of field-quadrature measurements},
  author = {Wiseman, H. M. and Milburn, G. J.},
  journal = {Phys. Rev. A},
  volume = {47},
  pages = {642},
  numpages = {0},
  year = {1993},
  publisher = {American Physical Society},
  doi = {10.1103/PhysRevA.47.642},
  url = {https://link.aps.org/doi/10.1103/PhysRevA.47.642}
}

@article{Nakajima_2016,
	doi = {10.1038/nphys3622},  
	url = {https://doi.org/10.1038%2Fnphys3622}, 
	year = 2016,
	volume = {12},
	pages = {296--300}, 
	author = {S. Nakajima and T. Tomita and S. Taie and Tomohiro Ichinose and Hideki Ozawa and Lei Wang and Matthias Troyer and Yoshiro Takahashi},  
	title = {Topological Thouless pumping of ultracold~fermions},
	journal = {Nat. Phys.}
}

@article{Haldane,
  title = {Model for a Quantum Hall Effect without Landau Levels: Condensed-Matter Realization of the "Parity Anomaly"},
  author = {Haldane, F. D. M.},
  journal = {Phys. Rev. Lett.},
  volume = {61},
  pages = {2015},
  year = {1988},
  publisher = {American Physical Society},
  doi = {10.1103/PhysRevLett.61.2015},
  url = {https://link.aps.org/doi/10.1103/PhysRevLett.61.2015}
}

@article{TKNN1982,
  title = {Quantized Hall Conductance in a Two-Dimensional Periodic Potential},
  author = {Thouless, D. J. and Kohmoto, M. and Nightingale, M. P. and den Nijs, M.},
  journal = {Phys. Rev. Lett.},
  volume = {49},
  %issue = {6},
  pages = {405},
  numpages = {0},
  year = {1982},
  %month= {08},
  publisher = {American Physical Society},
  doi = {10.1103/PhysRevLett.49.405},
  url = {https://link.aps.org/doi/10.1103/PhysRevLett.49.405}
}

@article{Thouless1983,
  title = {Quantization of particle transport},
  author = {Thouless, D. J.},
  journal = {Phys. Rev. B},
  volume = {27},
  %issue = {10},
  pages = {6083},
  numpages = {0},
  year = {1983},
  %month= {05},
  publisher = {American Physical Society},
  doi = {10.1103/PhysRevB.27.6083},
  url = {https://link.aps.org/doi/10.1103/PhysRevB.27.6083}
}

@article{PhysRevB.74.235111,
  title = {Insulator/Chern-insulator transition in the Haldane model},
  author = {Thonhauser, T. and Vanderbilt, D.},
  journal = {Phys. Rev. B},
  volume = {74},
  %issue = {23},
  pages = {235111},
  %numpages = {8},
  year = {2006},
  %%month= {12},
  publisher = {American Physical Society},
  doi = {10.1103/PhysRevB.74.235111},
  url = {https://link.aps.org/doi/10.1103/PhysRevB.74.235111}
}

@article{Ma2019,
  title = {{A dissipatively stabilized Mott insulator of photons}},
  author = {Ma, R. and Saxberg, B. and Owens, C. and Leung, N. and Lu, Y. and Simon, J. and Schuster, D. I.},
  journal = {Nature},
  volume = {566},
  pages = {51-57},
  year = {2019},
  publisher = {American Physical Society},
  doi = {10.1038/s41586-019-0897-9},
  url = {https://doi.org/10.1038/s41586-019-0897-9}
}

@article{Eckardt2017,
  title = {{Colloquium: Atomic quantum gases in periodically driven optical lattices}},
  author = {Eckardt, A.},
  journal = {Rev. Mod. Phys.},
  volume = {89},
  %issue = {1},
  pages = {011004},
  numpages = {30},
  year = {2017},
  %month= {},
  publisher = {American Physical Society},
  doi = {10.1103/RevModPhys.89.011004},
  url = {https://link.aps.org/doi/10.1103/RevModPhys.89.011004}
}

@article{2022arXiv220315670W,
   title={Quantum engineering of a synthetic thermal bath for bosonic atoms in a  one-dimensional optical lattice via Markovian feedback control},
   volume={13},
   ISSN={2542-4653},
   url={http://dx.doi.org/10.21468/SciPostPhys.13.3.059},
   DOI={10.21468/scipostphys.13.3.059},
   journal={SciPost Phys.},
   author={Wu, L.-N. and Eckardt, A.},
   year={2022} }

@Article{Barreiro2011,
author={Barreiro, Julio T.
and M{\"u}ller, Markus
and Schindler, Philipp
and Nigg, Daniel
and Monz, Thomas
and Chwalla, Michael
and Hennrich, Markus
and Roos, Christian F.
and Zoller, Peter
and Blatt, Rainer},
title={An open-system quantum simulator with trapped ions},
journal={Nature},
year={2011},
day={01},
volume={470},
pages={486-491},
doi={10.1038/nature09801},
url={https://doi.org/10.1038/nature09801}
}

@article{Sponselee_2018,
	doi = {10.1088/2058-9565/aadccd},
	url = {https://doi.org/10.1088%2F2058-9565%2Faadccd},
	year = 2018,
	%month= {sep},
	publisher = {{IOP} Publishing},
	volume = {4},
	pages = {014002},
	author = {K Sponselee and L Freystatzky and B Abeln and M Diem and B Hundt and A Kochanke and T Ponath and B Santra and L Mathey and K Sengstock and C Becker},
	title = {Dynamics of ultracold quantum gases in the dissipative Fermi{\textendash}Hubbard model},
	journal = {Quantum Sci. Technol.},
	abstract = {We employ metastable ultracold 173Yb atoms to study dynamics in the 1D dissipative Fermi–Hubbard model experimentally and theoretically, and observe a complete inhibition of two-body losses after initial fast transient dynamics. We attribute the suppression of particle loss to the dynamical generation of a highly entangled Dicke state. For several lattice depths and for two- and six-spin component mixtures we find very similar dynamics, showing that the creation of strongly correlated states is a robust and universal phenomenon. This offers interesting opportunities for precision measurements.}
}

@article {Tomitae1701513,
	author = {Tomita, Takafumi and Nakajima, Shuta and Danshita, Ippei and Takasu, Yosuke and Takahashi, Yoshiro},
	title = {Observation of the Mott insulator to superfluid crossover of a driven-dissipative Bose-Hubbard system},
	volume = {3},
	year = {2017},
    pages = {e1701513},
	doi = {10.1126/sciadv.1701513},
	publisher = {American Association for the Advancement of Science},
	URL = {https://doi.org/10.1126/sciadv.1701513},
	journal = {Sci. Adv.}
}

@article{Liu_2021,
	doi = {10.1103/physrevresearch.3.043119},
	url = {https://doi.org/10.1103%2Fphysrevresearch.3.043119},
	year = {2021}, 
    pages = {043119},
	volume = {3},  
	author = {Z. Liu and E.~J. Bergholtz and J.~C. Budich},  
	title = {Dissipative preparation of fractional Chern insulators},  
	journal = {Phys. Rev. Res.}
}

@article{PhysRevLett.42.1698,
  title = {Solitons in Polyacetylene},
  author = {Su, W. P. and Schrieffer, J. R. and Heeger, A. J.},
  journal = {Phys. Rev. Lett.},
  volume = {42},
  %issue = {25},
  pages = {1698},

  %numpages = {0},
  year = {1979},
  %%month= {06},
  publisher = {American Physical Society},
  doi = {10.1103/PhysRevLett.42.1698},
  url = {https://link.aps.org/doi/10.1103/PhysRevLett.42.1698}
}

@book{Bernevig2013,
	publisher = {Princeton University Press},
	title = {Topological Insulators and Topological Superconductors},
	author = {Bernevig, B.~A.},
	year = {2013}
}

@Article{Goldstein2019,
	title={{Dissipation-induced topological insulators: A no-go theorem and a recipe}},
	author={M. Goldstein},
	journal={SciPost Phys.},
	volume={7},
	%issue={5},
	pages={67},
	year={2019},
	publisher={SciPost},
	doi={10.21468/SciPostPhys.7.5.067},
	url={https://scipost.org/10.21468/SciPostPhys.7.5.067},
}

@article{Kitagawa2010,
  title = {Topological characterization of periodically driven quantum systems},
  author = {Kitagawa, T. and Berg, E. and Rudner, M. and Demler, E.},
  journal = {Phys. Rev. B},
  volume = {82},
  %issue = {23},
  pages = {235114},
  numpages = {12},
  year = {2010},
  %month= {12},
  publisher = {American Physical Society},
  doi = {10.1103/PhysRevB.82.235114},
  url = {https://link.aps.org/doi/10.1103/PhysRevB.82.235114}
}

@Article{Diehl2011,
author={Diehl, S.
and Rico, E.
and Baranov, M.~A.
and Zoller, P.},
title={Topology by dissipation in atomic quantum wires},
journal={Nat. Phys.},
year={2011},
day={01},
volume={7},
pages={971-977},
issn={1745-2481},
doi={10.1038/nphys2106},
url={https://doi.org/10.1038/nphys2106}
}

@article{PhysRevB.102.184302,
  title = {Dissipative preparation of many-body Floquet Chern insulators},
  author = {Bandyopadhyay, S. and Dutta, A.},
  journal = {Phys. Rev. B},
  volume = {102},
  %issue = {18},
  pages = {184302},
  numpages = {17},
  year = {2020},
  %month= {11},
  publisher = {American Physical Society},
  doi = {10.1103/PhysRevB.102.184302},
  url = {https://link.aps.org/doi/10.1103/PhysRevB.102.184302}
}

@Article{Goldman2016,
author={Goldman, N.
and Budich, J. C.
and Zoller, P.},
title={Topological quantum matter with ultracold gases in optical lattices},
journal={Nat. Phys.},
year={2016},
day={01},
volume={12},
pages={639-645},
issn={1745-2481},
doi={10.1038/nphys3803},
url={https://doi.org/10.1038/nphys3803}
}

@Article{Verstraete2009,
author={Verstraete, F.
and Wolf, M.~M.
and Cirac, J.~I.},
title={Quantum computation and quantum-state engineering driven by dissipation},
journal={Nat. Phys.},
year={2009},
day={01},
volume={5},
pages={633-636},
issn={1745-2481},
doi={10.1038/nphys1342},
url={https://doi.org/10.1038/nphys1342}
}

@Article{Diehl2008,
author={Diehl, S.
and Micheli, A.
and Kantian, A.
and Kraus, B.
and B{\"u}chler, H. P.
and Zoller, P.},
title={Quantum states and phases in driven open quantum systems with cold atoms},
journal={Nat. Phys.},
year={2008},
volume={4},
pages={878-883},
issn={1745-2481},
doi={10.1038/nphys1073},
url={https://doi.org/10.1038/nphys1073}
}

@article{Nakajima_2021,
	doi = {10.1038/s41567-021-01229-9},  
	url = {https://doi.org/10.1038%2Fs41567-021-01229-9},  
	year = 2021,
	%%month= {apr},  
	publisher = {Springer Science and Business Media {LLC}},  
	volume = {17},  
	pages = {844--849},
	author = {Shuta Nakajima and Nobuyuki Takei and Keita Sakuma and Yoshihito Kuno and Pasquale Marra and Yoshiro Takahashi}, 
	title = {Competition and interplay between topology and quasi-periodic disorder in Thouless pumping of ultracold atoms}, 
	journal = {Nat. Phys.}
}

@article{PhysRevLett.115.095301,
  title = {Diffraction-Unlimited Position Measurement of Ultracold Atoms in an Optical Lattice},
  author = {Ashida, Y. and Ueda, M.},
  journal = {Phys. Rev. Lett.},
  volume = {115},
  %issue = {9},
  pages = {095301},
  year = {2015},
  %month= {08},
  publisher = {American Physical Society},
  doi = {10.1103/PhysRevLett.115.095301},
  url = {https://link.aps.org/doi/10.1103/PhysRevLett.115.095301}
}

@article{PhysRevLett.114.113604,
  title = {Multipartite Entangled Spatial Modes of Ultracold Atoms Generated and Controlled by Quantum Measurement},
  author = {Elliott, T. J. and Kozlowski, W. and Caballero-Benitez, S. F. and Mekhov, I. B.},
  journal = {Phys. Rev. Lett.},
  volume = {114},
  %issue = {11},
  pages = {113604},
  %numpages = {5},
  year = {2015},
  %%month= {},
  publisher = {American Physical Society},
  doi = {10.1103/PhysRevLett.114.113604},
  url = {https://link.aps.org/doi/10.1103/PhysRevLett.114.113604}
}

@article{RevModPhys.85.553,
  title = {Cold atoms in cavity-generated dynamical optical potentials},
  author = {Ritsch, H. and Domokos, P. and Brennecke, F. and Esslinger, T.},
  journal = {Rev. Mod. Phys.},
  volume = {85},
  %issue = {2},
  pages = {553},
  %numpages = {0},
  year = {2013},
  %%month= {04},
  publisher = {American Physical Society},
  doi = {10.1103/RevModPhys.85.553},
  url = {https://link.aps.org/doi/10.1103/RevModPhys.85.553}
}

\nolinenumbers

\end{document}